\documentclass{article}

\usepackage{amsmath}
\usepackage{amsthm}
\usepackage{amsfonts}
\usepackage{amssymb}
\usepackage{graphicx, rotating}
\usepackage{epstopdf}
\usepackage{epsfig}
\usepackage{latexsym}
\usepackage{graphicx}
\usepackage{color}
\usepackage[dvipsnames]{xcolor}
\usepackage{amsmath,amssymb}
\usepackage{cite}
\usepackage{slashed}
\usepackage{float}
\usepackage[export]{adjustbox}
\usepackage{physics}
\usepackage{textgreek}
\usepackage[dvipsnames]{xcolor}
\definecolor{rossos}{cmyk}{0,1,1,0.55}
\definecolor{bluscuro}{rgb}{0.15, 0.2, .85}
\definecolor{bluchiaro}{cmyk}{1,.3,0.,0.1}
\usepackage[colorlinks=True, citecolor=bluscuro, linkcolor=black, urlcolor=bluscuro,linktocpage]{hyperref}
\usepackage{orcidlink}
\usepackage{authblk}

\def\S{\text{\tiny S}}

\begin{document}
	
	\title{Shadow Sectors of Gauge Theories}
	\author[1]{Loris Del Grosso\hspace{0.05cm}\orcidlink{0000-0002-6722-4629}\thanks{\href{mailto:ldelgro1@jh.edu}{ldelgro1@jh.edu}}}
	\author[1,2]{David E. Kaplan\hspace{0.05cm}\orcidlink{https://orcid.org/0000-0001-8175-4506}\thanks{\href{mailto:david.kaplan@jhu.edu}{david.kaplan@jhu.edu}}}
	\author[1]{Francesco Serra\hspace{0.05cm}\orcidlink{0000-0002-7962-3555}\thanks{\href{mailto:fserra2@jh.edu}{fserra2@jh.edu}}}
	
	\affil[1]{\small Department of Physics \& Astronomy, The Johns Hopkins University, Baltimore, MD 21218, USA}
	\affil[2]{\small Kavli IPMU (WPI), UTIAS, The University of Tokyo, Kashiwa, Chiba 277-8583, Japan}
	
	\date{\today}
	
	\maketitle
	
	\begin{abstract}
		\noindent
		We show that both abelian and non-abelian gauge theories admit configurations in which the fields behave as if in the presence of static charge densities, or ``shadow charges". These correspond to nontrivial initial conditions for the fields that generate gauge transformations, the Gauss' law operators. 
        In non-abelian theories, such configurations seem to demand additional physical fields with exactly static charge densities. In contrast with this expectation, we show that gauge theory alone provides a consistent and gauge-invariant description of shadow charges. Canonical quantization then yields continuous shadow charges for abelian theories and quantized ones for non-abelian theories. In general, our findings indicate that all local conservation laws give rise to gauge symmetries, even in the presence of second-class constraints.
        		
	\end{abstract}

	\tableofcontents
	\newpage
	\section{Introduction}\label{sec:intro}
Many fundamental interactions found in nature are mediated by gapless spinning bosons. Notably, Lorentz invariance implies that such interactions are independent of arbitrary changes in the longitudinal components of the mediators \cite{Weinberg:1964ew,Weinberg:1965rz}. This redundancy is called a gauge symmetry, and it implies universal features of nature. For instance, in electromagnetism, gauge symmetry enforces conservation of the electric charge; in gravity, it results in the equivalence principle. 
While Lorentz invariance is an important guiding principle for phenomenology, the fundamental basis of gauge symmetry lies in local conservation laws that are not tied to relativity and hold even when Lorentz symmetry is not manifest.
 
In electromagnetism, for instance, gauge symmetry can be understood as arising from a degeneracy in the Hamiltonian of the system:  the Hamiltonian does not depend on the longitudinal component of the vector potential. By Hamilton's equations, this degeneracy implies that the conjugate momentum of the longitudinal component of the field has a vanishing time derivative -- a conservation law. When the field is coupled to a conserved current $J^\mu$, this Hamilton's equation becomes:
\begin{align}\label{eq:gaussconst}
  \frac{d}{d t}(\vec{\nabla}\!\cdot\!\vec{E}-J^0)=0\;,
\end{align}
where $\vec{E}$ is the electric field and $J^0$ is the electric charge density of matter. This conservation law directly encodes the extent to which gauge symmetry restricts the dynamics of the system.
This equation can be trivially integrated by choosing an integration constant, an initial condition for the system. Historically, such integration constant has been set to zero. This leads to a theory with manifest Lorentz symmetry, in which the time component of the vector potential, $A_0$, can be introduced as a Lagrange multiplier for Gauss' law: $\vec{\nabla}\!\cdot\!\vec{E}-J^0=0$.
However, the full phase space of the theory contains a broader set of configurations, in which the integration constant is not set to zero, and the fields respond to non-trivial initial conditions:
$
  \vec{\nabla}\!\cdot\!\vec{E}-J^0=\rho(\vec{x})\;,
$
where $\rho(\vec{x})$ an arbitrary time-independent function, encoding the possible initial conditions for the field configurations \cite{Kaplan:2023fbl,Kaplan:2023wyw,DelGrosso:2024gnj}. Due to its appearance as a constant source term in  Gauss' law with no matter counterpart, we call $\rho(\vec{x})$ a shadow charge density. The full phase space of the theory is foliated into regions corresponding to different choices of $\rho(\vec{x})$, the so-called constraint surfaces.

In analogy with this result, one finds that in classical Yang-Mills theory the non-abelian Gauss' law operators $G_a=\vec{D}\!\cdot\!\vec{E}_a$ are locally conserved: \begin{align}\frac{d}{dt}G_a=0\;.\end{align} 
The theory therefore suggests the existence of configurations with non-trivial initial conditions for these constrained quantities, {\it i.e.}, non-abelian shadow charges: $G_a=\rho_a(\vec{x})$. However, already at the level of Poisson brackets, the non-abelian Gauss' law operators do not commute with each other, meaning that some of these initial conditions will be gauge-dependent for non-vanishing shadow charge. This becomes a problem when one quantizes the theory, as it is difficult to impose such initial conditions, e.g. in expectation values, while removing gauge redundancies from the Hilbert space.
In Dirac's terminology, the constraints of the theory have become second-class, as opposed to gauge-invariant, first-class constraints \cite{Dirac:1950pj,Henneaux:1992ig}.
Traditionally, second-class constraints have been regarded as an obstruction to gauge symmetry, which can only be restored if new physical, static fields are added to the theory, see e.g. \cite{Batalin:1986aq,Batalin:1986fm,Batalin:1991jm}. 
This would suggest that shadow charges can only be quantized consistently if the theory is coupled to new degrees of freedom with infinite inertial mass. 

In the present work, contrary to this view, we show that the second-class constraints themselves generate genuine gauge transformations in a region of phase space defined by the subset of gauge invariant initial conditions, which we call the constraint orbit. This is the union of all gauge-equivalent constraint surfaces of the non-abelian theory. This approach provides a gauge-invariant description of shadow charges without introducing physical fields with infinite mass. We can then consistently quantize the pure gauge theory at non-zero shadow charge, obtaining well-defined matrix elements in a non-degenerate Hilbert space. 

The work is organized as follows.
We begin by analyzing the classical phase space of abelian and non-abelian gauge theory in Sec.~\ref{sec:classical}, showing that configurations with non-vanishing shadow charge arise naturally and remain fully compatible with gauge symmetry. In the non-abelian case, we discuss the standard treatment of second-class constraints and we introduce our proposal based on the constraint orbit.
The algebra of gauge-invariant observables is then constructed in Sec.~\ref{sec:gauge} using the canonical BRST formalism \cite{Becchi:1974md,Becchi:1974xu,Tyutin:1975qk,Becchi:1975nq,Barnich:2000zw,Henneaux:1985kr,Fradkin:1975cq,Batalin:1977pb,Batalin:1981jr,Batalin:1983pz,Batalin:1985qj,Batalin:1989dm,Batalin:1983ggl,Anderson:1951ta}. In particular, we extend the BRST construction to the presence of shadow charges and to the constraint-orbit picture, and discuss gauge fixing.
With these classical results, we carry out quantization in Sec.~\ref{sec:quantum}. Canonical quantization promotes the classical BRST charge to the nilpotent BRST operator of the quantum theory, defining a non-degenerate physical Hilbert space where gauge transformations act trivially \cite{Becchi:1974md,Becchi:1974xu,Becchi:1975nq,Tyutin:1975qk,Kugo:1977yx,Kugo:1977zq,Kugo:1979gm}. In QED, we show explicitly that unphysical excitations decouple through the quartet mechanism and that states with definite shadow charge can be consistently constructed. In Yang-Mills theory, we demonstrate that our constraint-orbit formulation yields the same gauge-invariant matrix elements as the traditional approach with static fields, and we find that the non-abelian shadow charges are quantized. Finally, in Sec.~\ref{sec:conclusions} we summarize the main results.
In the Appendices we provide additional detail on various technical aspects of the discussion: Dirac Brackets, App.~\ref{app:rigidbody}, second-class constraints and additional fields, App.~\ref{sec:rhocanonical}, algebraic resolution of a constraint surface, App.~\ref{app:resolution}, gauge symmetry in the non-abelian case, App.~\ref{app:gaugesymm}, boosted shadow charges, App.~\ref{sec:boosts}.

	\section{Shadow charges in classical gauge theory}\label{sec:classical}
	In this section, we show that the phase space of gauge theories contains configurations in which the fields follow static charge densities that do not correspond to any charged degrees of freedom, the shadow charges.
 These shadow charges correspond to non-trivial initial conditions for the electric field, which remain imprinted in the field configuration at all times.
 We characterize the admissible shadow charge densities, and we show how gauge symmetry is a feature of all of these configurations. 
 
 We start by studying the phase space of electromagnetism. We then move to the case of Yang-Mills theories.
 
 \subsection{Shadow charges in Electromagnetism}\label{sec:clElectro}
    We start by showing how the phase space describing electromagnetic photons contains a wider set of configurations compared to what is derived from the full set of Maxwell's equations. As mentioned, this is because Gauss' law involves a specific choice of initial conditions. Non-trivial initial conditions lead to new configurations, which are however, characterized by the same dynamics and the same degrees of freedom, regardless of the initial conditions.
 In order to begin, we first have to decide what the canonical variables describing the photon are. For instance, we could start using only the spatial vector field $\vec{A}$, and its conjugate momentum $\vec{\Pi}$, with canonical Poisson brackets. For simplicity of notation, in the following, we will write Poisson brackets omitting spatial Dirac delta functions. In order to make closer contact with the Lorentz invariant formulation of the theory, we can then add the time component of the vector field, $A_0$, and its conjugate momentum $\Pi_0$. As we will see explicitly in the rest of the work, see e.g. Sec.~\ref{sec:gaugefixing}, the resulting dynamics is independent on this choice, as $A_0$ is in any case a gauge variable. 
 As a matter of fact, we find it useful to describe the dynamics using the following Hamiltonian, independent on $A_0$:
 \begin{align}\label{eq:H}
  H=\int d^3 x\,\Bigg(\frac{1}{2}\vec{\Pi}^2+\frac{1}{2}(\vec{\nabla}\!\wedge\!\vec{A})^2\Bigg)\;,
 \end{align}
 As we will discuss, this choice of Hamiltonian is equivalent to many others that can be obtained from this by gauge fixing. For instance, we will see that gauge fixing can make the Hamiltonian depend on $A_0$. However, as we will discuss in detail in Sec.~\ref{sec:gaugefixing},  Eq.~\eqref{eq:H} is the only form of the Hamiltonian to be gauge invariant across the whole phase space.
 For now, we can limit ourselves to notice that this Hamiltonian leads to Ampere's law and Faraday's law for the electric field $\vec{\Pi}$ and the magnetic field $\vec{\nabla}\wedge\vec{A}$:
 \begin{align}\label{eq:ampere}
  &\frac{d}{dt}\vec{\Pi}=\{\vec{\Pi},H\}=\vec{\nabla}\wedge(\vec{\nabla}\wedge\vec{A})\;,\\
  &\frac{d}{dt}\vec{\nabla}\wedge\vec{A}=\{\vec{\nabla}\wedge\vec{A},H\}=\vec{\nabla}\wedge\vec{\Pi}\;,
 \end{align}
 confirming that $H$ describes electrodynamics well.
 The right-hand side of Eq.~\eqref{eq:ampere} has zero divergence. We can understand the reason for this by noting that this Hamiltonian is independent of the longitudinal part of $\vec{A}$. As a result, Hamilton's equations imply that the longitudinal component of $\Pi$ has zero time derivative, which is locally expressed as:
 \begin{align}\label{eq:dyn}
  \frac{d}{dt}\vec{\nabla}\!\cdot\!\vec{\Pi}=0\;.
 \end{align}
 This equation is part of Ampere's law, and it implies a constant value for $\vec{\nabla}\!\cdot\!\vec{\Pi}$.
 A simple choice is the following:
 \begin{align}\label{eq:nondyn}
  \vec{\nabla}\!\cdot\!\vec{\Pi}\approx 0\;,
 \end{align}
 where the weak-equality symbol $\approx$ indicates that the equality follows from our choice of initial conditions for the electric field, rather than from algebraic properties that hold for every configuration. As a matter of fact, Eq.~\eqref{eq:dyn} has other solutions besides Eq.~\eqref{eq:nondyn}.
 
 Before investigating these additional solutions, we characterize Eq.~\eqref{eq:nondyn} in a canonical way. Since time evolution does not change the values of Eq.~\eqref{eq:nondyn}, this equation defines a surface in phase space that remains unchanged under time evolution. This surface is called a constraint surface, as on this surface the (apparently) dynamical equation \eqref{eq:dyn} can be traded for a non-dynamical equation \eqref{eq:nondyn}, a specific solution. This effectively reduces -- or constrains -- the apparent dynamics of the system.
 For this reason, we call Eq.~\eqref{eq:nondyn} a constraint of the system, which we indicate as:
 \begin{align}\label{eq:constr0}
  \phi_1=\vec{\nabla}\!\cdot\!\vec{\Pi}\approx 0\;.
 \end{align}
 Note that in general, the constraint surface can only be invariant under time evolution if all the time derivatives of the constraints vanish on the constraint surface. In this case, this is due to the independence of $H$ on the longitudinal part of $\vec{A}$.
 
 Since the Hamiltonian $H$ is independent of the canonical pair $(A_0\,,\,\Pi_0)$, these quantities will also be constant in time. Therefore, the constraint surface can be extended in a Lorentz-friendly way by adding the constraint:
 \begin{align}
  \phi_0=\Pi_0\approx 0\;,
 \end{align}
 which is a solution to the Hamilton's equation $\dot{\Pi}_0=0$.
 
 Besides characterizing the phase space and the physical configurations of the theory, the constraints are the canonical generators of gauge transformations on a given constraint surface. 
 In fact, the time translation invariance of the constraint surface stems from the independence of the Hamiltonian on the conjugate variables to the constraints, $A_0$, and the longitudinal part of $\vec{A}$ in our case. This degeneracy at the level of the Hamiltonian means that the variables that are conjugate to the constraints can be arbitrarily shifted without changing the predictions of the theory, which is called a gauge symmetry.
 These shifts, called gauge transformations, are generated in phase space by taking the Poisson brackets of any quantity with the constraints, multiplied by arbitrary functions of space:
 \begin{align}
  \delta_u F=u_0(\vec{x})\{F,\phi_0\}+u_1(\vec{x})\{F,\phi_1\}\;.
 \end{align}
    These functions need to be time independent in order for the transformations to be compatible with Hamilton's equations: $\dot{A}_0=0\,,\,\dot{\vec{A}}=\vec{\Pi}$. 
 Importantly, since the initial condition for the electric field is a physical observable, $\phi_1$ must also be gauge invariant. This is granted as the constraints have vanishing Poisson brackets with each other:
 \begin{align}\label{eq:firstclass}
  \{\phi_0,\phi_1\}\approx 0.
 \end{align}
 Constraints of this kind are called first-class, as opposed to constraints that have non-vanishing Poisson brackets on the constraint surface, which are called second-class. Usually, one extends the requirement of Eq.~\eqref{eq:firstclass} by asking for the whole constraint surface to be invariant under gauge transformations. This requirement, trivially satisfied in the present case, restricts gauge transformations to those generated by first-class constraints. We will discuss this choice more carefully in Sec.~\ref{sec:constraintorbit}, where we will address the role of second-class constraints in the phase space of Yang-Mills theory.
 
 Having defined gauge transformations in phase space, it is simple to see how these act on the canonical variables:
 \begin{align}\label{eq:gaugetransf}
  \delta_u\vec{A}=\vec{\nabla}u_1\;,\quad\delta_u A_0=u_0\;,\quad\delta_u\vec{\Pi}=\delta_u\Pi_0=0\;.
 \end{align}
 As we will discuss in Sec.~\ref{sec:gaugefixing}, these transformations are mapped to the usual gauge transformations $A_\mu\to A_\mu+\partial_\mu\lambda$, when one introduces a dependence on $A_0$ in the Hamiltonian. 
 
 Having presented the canonical description of the gauge redundancies, we now proceed to describe the rest of the phase space of electromagnetism. This is captured by choosing other solutions to Eq.~\eqref{eq:dyn}, \textit{i.e.}, initial conditions that differ from Eq.~\eqref{eq:nondyn}. (We could extend our analysis to different solutions for $\Pi_0$, but this change would not affect any of the predictions of the theory, as the Hamiltonian is independent of $\Pi_0$.) For instance, we can solve Eq.~\eqref{eq:dyn} by choosing:
 \begin{align}
  \vec{\nabla}\!\cdot\!\vec{\Pi}\approx\rho(\vec{x})\;,
 \end{align}
 This choice maps out a new set of configurations in which Gauss' law is modified by a non-vanishing shadow charge density $\rho(\vec{x})$, constant in time. The shadow charge does not correspond to additional charged degrees of freedom. Instead, it encodes permanent features of the electric field configuration. These configurations are allowed by the dynamics and characterize the full phase space of the theory.
 
 Furthermore, our choice of shadow charge singles out a region of phase space that is mapped into itself under time evolution -- a new constraint surface. This constraint surface will be given by:
 \begin{align}\label{eq:constrrho}
  \phi_0=\Pi_0\approx 0\;,\quad\phi_1=\vec{\nabla}\!\cdot\!\vec{\Pi}-\rho(\vec{x})\approx 0\;.
 \end{align}
 Since the shadow charge is just a c-number function, independent of the canonical variables, its Poisson brackets vanish with any other quantity. This means that the Poisson brackets between the two constraints $\phi_0$ and $\phi_1$ are vanishing regardless of $\rho$, and these two quantities generate the same gauge transformations discussed before.
 As before, both the Hamiltonian and the constraint surface are invariant under these gauge transformations, confirming that the shadow charge is a physical and (if there were charged matter in the theory) observable parameter.
 As we will show in Sec.~\ref{sec:gaugefixing}, these configurations can be described from the point of view of the action, by performing a gauge fixing of the Hamiltonian and a Legendre transformation. While the Hamiltonian $H$ describes the dynamics across the whole phase space, the action will depend on the choice of constraint surface.
 
 From this discussion, we conclude that the theory of electromagnetism has configurations characterized by non-vanishing constant shadow charge densities, which foliate the phase space. These shadow charges reflect the freedom in initial conditions for the electric field. Moreover, we have shown that these configurations have gauge redundancies, regardless of their shadow charge. Both the dynamics and the redundancies are captured by the Hamiltonian $H$ throughout the whole phase space, in a way that is consistent with the usual formulation of electromagnetism.
 
 While these shadow charges are all constant in time, each constraint surface contains configurations in which all of the field excitations are boosted with respect to the shadow charge. This means that physical excitations can observe classical shadow charges with arbitrary boosts.
 While these considerations indicate that shadow charges are compatible with Lorentz invariance as a fundamental feature, the canonical point of view leaves open the possibility for the theory to be formulated without introducing $A_0$ and $\Pi_0$, and in a fundamentally Lorentz-breaking way.
 
 Moving on, we study the case of Yang-Mills theory and characterize its full phase space in terms of non-abelian shadow charges.

 \subsection{Shadow charges in Yang-Mills theory}\label{sec:clYM}
    Having discussed the presence of shadow charges in classical electromagnetism, we now study the case of Yang-Mills theory.
 Following the steps of Sec.~\ref{sec:clElectro}, we first examine the configuration with a trivial initial condition for the Gauss' law operators, and then we extend our approach to the rest of the phase space.
 
 Similarly to before, we can start with the canonical variables $A_{ai}\,,\,\Pi_{ai}$, with $a=1,\dots, N^2-1$, and we are free to include the temporal components of the vector fields and their conjugates: $A_{a0}\,,\,\Pi_{a0}$. Whether the temporal components are considered fundamental variables or not makes little difference, as the $A_{a0}$ always turn out to be gauge redundancies.
 Similarly to before, we choose to describe the dynamics of gluons with a Hamiltonian that does not depend on the $A_{a0}$:
 \begin{align}\label{eq:HYM}
  H=\int d^3x\;\Big(\frac{1}{2}\Pi_{ai}\Pi_{ai}+\frac{1}{4}F_{aij}F_{aij}\Big)\;,
 \end{align}
 with field strength defined as:
 \begin{align}
  F_{a\mu\nu}=\partial_{\mu}A_{a\nu}-\partial_{\nu}A_{a\mu}+C_{abc}A_{b\mu}A_{c\nu}\;,
 \end{align}
 with $C_{abc}$ the structure constants of $SU(N)$, and the self-interaction coupling set to $g=1$ for simplicity of notation.
 As in electromagnetism, this is the only Hamiltonian to be gauge invariant across the whole phase space of the theory. For a fixed configuration, this Hamiltonian can be consistently gauge-fixed in a great number of ways, which allows one to derive more familiar forms of the Hamiltonian and of the action with explicit dependence on e.g. the $A_{a0}$. We will show this in Sec.~\ref{sec:gaugefixing}.
 For the moment, we can limit ourselves to seeing that our choice reproduces the non-abelian version of Ampere's law:
 \begin{align}\label{eq:ampereYM}
  \frac{d}{dt}\Pi_{ai}=\{\Pi_{ai},H\}=\partial_jF_{aij}+C_{abc}A_{bj}F_{cij}\equiv (D_jF_{ij})_a\;,
 \end{align}
 where the last equality defines the covariant derivative.
 Similarly to the Abelian case, the antisymmetry of $F_{aij}$ implies the conservation of the covariant divergence of $\Pi_{ai}$, as one of Hamilton's equations. We call these quantities the Gauss' law operators, $G_a$:
 \begin{align}
  G_a=\partial_i\Pi_{ai}+C_{abc}A_{bi}\Pi_{ci}=D_i\Pi_{ai}\;.
 \end{align}
 To see that these are constant, we can take the covariant divergence of the non-abelian Ampere's law:
 \begin{align}
  D_i\dot{\Pi}_{ai}=\partial_i\dot{\Pi}_{ai}+C_{abc}A_{bi}\dot{\Pi}_{ci}=\frac{d}{dt}(\partial_i\Pi_{ai}+C_{abc}A_{bi}\Pi_{ci})-C_{abc}\dot{A}_{bi}\Pi_{ci}\;.
 \end{align}
 Using the Hamilton equation for $A_{ai}$, $\dot{A}_{ai}=\Pi_{ai}$, we see that the last term vanishes and we obtain:
 \begin{align}\label{eq:gaussdot}
  &\dot{G}_a=\{G_a,H\}=D_i\dot{\Pi}_{ai}=(D_iD_jF_{ij})_a=0\;.
 \end{align} 
 This conservation law is the non-abelian equivalent of Eq.~\eqref{eq:dyn}.
 Since again the Hamiltonian has no dependence on the canonical pairs $(A_{a0}\,,\,\Pi_{a0})$, we will have $\dot{\Pi}_{a0}=\{\Pi_{a0},H\}=0$, meaning that the phase space can be described in terms of constraint surfaces described by the initial conditions for $\Pi_0$ and the $G_a$. For instance, picking trivial initial conditions, we have the following constraints:
 \begin{align}\label{eq:constr0YM}
  \phi_{0a}=\Pi_{a0}\approx 0\;,\quad\phi_{1a}=G_a\approx 0\;.
 \end{align}
 These equations define a constraint surface invariant under time evolution, as in the case of electromagnetism. Clearly, the Hamiltonian is invariant under the transformations generated by these constraints: $\{H,\phi_{0a}\}=\{H,\phi_{1a}\}=0$. These are therefore gauge transformations -- no prediction depends on whether the fields have been transformed by acting with the constraints. Compatibly with the common requirement, the constraint surface itself is invariant under these transformations. Indeed, the only non-trivial Poisson brackets among the constraints are the following:
 \begin{align}\label{eq:gausscommut}
  \{\phi_{1a},\phi_{1b}\}=\{G_a,G_b\}=C_{abc}G_c\;,
 \end{align}
 which vanish on the constraint surface of Eq.~\eqref{eq:constr0YM}, $C_{abc}G_c=C_{abc}\phi_{1c}\approx 0$. In other words, the constraints are first-class for this choice of initial conditions.
 These constraints act on the canonical variables as follows:
 \begin{align}\label{eq:gaugetransfYM}
  u_{1b}\{A_{ai},G_b\}=\partial_i u_{1a}+C_{abc}A_{bi}u_{1c}=(D_iu_1)_a\;,\quad u_{0b}\{A_{a0},\Pi_{b0}\}=u_{0a}\;.
 \end{align}
 Similarly to the case of electromagnetism, we will see that these gauge transformations are mapped in the usual ones, $A_{a\mu}\to A_{a\mu}+(D_\mu\lambda)_a$, when the Hamiltonian is gauge-fixed so as to have an explicit dependence on $A_{a0}$, see Sec.~\ref{sec:gaugefixing} and App.~\ref{app:gaugesymm}.
 In analogy with electromagnetism, we now can see that different solutions to Eq.~\eqref{eq:gaussdot} describe the rest of the phase space of the theory.
 In fact, we can take a solution of the form:
 \begin{align}
  G_a\approx \rho_a(\vec{x})\;.
 \end{align}
 Which means that on these configurations the field configuration follows a time-independent shadow charge density, described by the $\rho_a(\vec{x})$.
 Again, since the Hamiltonian is independent of the $\Pi_{a0}$, we do not explore non-trivial choices for their values.
 These new configurations will be associated with new constraint surfaces:
 \begin{align}\label{eq:constrYM}
  \phi_{0a}=\Pi_{a0}\approx0\;,\quad\phi_{1a}=G_a-\rho_a\approx 0\;.
 \end{align}
 Similarly to the previous cases, the constraints generate transformations that leave the Hamiltonian invariant. Therefore, the shifts of the fields under these transformations naturally fit the role of a gauge symmetry of the theory. Moreover, since the $\rho_a$ are c-number functions, they have vanishing Poisson brackets with every quantity, meaning that the $\phi_{ia}$ generate the same transformations as in Eq.~\eqref{eq:gaugetransfYM}.
 
 Despite these similarities with the case of zero shadow charges, the presence of non-vanishing shadow charges $\rho_a$ leads to an important difference. That is, the constraints $\phi_{1a}$ no longer have vanishing Poisson brackets:
 \begin{align}\label{eq:secondclassYM}
  \{\phi_{1a},\phi_{1b}\}=\{G_a-\rho_a,G_b-\rho_b\}=C_{abc}G_c\approx C_{abc}\rho_c\neq 0\;.
 \end{align}
 In other words, the constraints are second-class.
 This means that the constraint surface is no longer mapped into itself by all of the transformations generated by the constraints.
 
 Historically, transformations generated by second-class constraints have not been considered generators of gauge symmetries, since they change the constraint surface, \textit{i.e.}, the initial conditions of the constrained fields. 

    For instance, a commonly proposed way to deal with second-class constraints is to eliminate them from the dynamics, see e.g. \cite{Dirac:1950pj}.
 Ideally, one might want to do this by simply solving the constraints and plugging back the solutions into the remaining Hamilton's equations. When this is not possible in a configuration-independent way, one can achieve the elimination of the second-class constraints by changing the bracket algebra of phase space in such a way that the second-class constraints become trivial. In practice, this is obtained by replacing the canonical Poisson brackets with the so-called Dirac brackets. These brackets have the effect of demoting the second-class constraints to functions that commute with every quantity, discarding the physical transformations that the second-class constraints generate with respect to the Poisson brackets.
 In practice, given constraints with Poisson brackets $\{\phi_\alpha,\phi_\beta\}\approx \mathcal{B}_{\alpha\beta}\neq 0$, one defines the Dirac brackets as:
 \begin{align}\begin{split}\label{eq:diracbrack}
   &\{A,B\}_D=\{A,B\}-\{A,\phi_\alpha\}\mathcal{B}^{-1}_{\alpha\beta}\{\phi_\beta,B\}\;,\\
   &\text{leading to:}\quad\{\phi_\alpha,F\}_D=0\;,\quad \text{for all second class}\;\phi_\alpha \;\text{and for all}\;F,\end{split}
 \end{align}
 where the inverse of $\mathcal{B}$ is defined on the maximal subspace over which $\mathcal{B}$ is invertible.
 As we show in App.~\ref{app:rigidbody} in the context of the rigid body, using Dirac brackets is equivalent to studying the dynamics up to precession effects. Since second-class constraints act trivially in terms of Dirac brackets, from this point of view, it appears that second-class constraints do not generate gauge symmetry, but rather break it. As we will see, this is not the case, and second-class constraints are in fact associated to gauge transformations.
 The Dirac brackets approach is often impractical when one quantizes the theory, as it is difficult to find a set of operators whose commutation relations reproduce the Dirac brackets algebra, see e.g. Sec.~\ref{sec:QYM}. Two other approaches similar in spirit have been discussed in \cite{Faddeev:1988qp,Vytheeswaran:1994np,Anishetty:1992yk,Vytheeswaran:1999da}.
    Contrary to this approach, in the following, we describe two ways to study a system with second-class constraints, in a way that avoids the problems of Dirac's brackets and makes symmetry manifest.
    First, in Sec.~\ref{sec:UVfields} we will describe a known way to do so, which is enlarging the system by adding new physical canonical variables and changing the constraints so as to make them first class, see e.g. \cite{Henneaux:1992ig,Batalin:1986aq,Batalin:1986fm,Batalin:1991jm}. This approach relies on introducing new physical fields which, as we will discuss, should be understood in terms of non-standard UV physics.
    Second, in Sec.~\ref{sec:constraintorbit}, we will introduce a new approach to retrieve gauge symmetry from second-class constraints without altering the theory and without introducing additional UV fields. In this approach, shadow charges are consistently described as IR features of the theory. 
    
    \subsubsection{Shadow charges as problematic UV fields?}
    \label{sec:UVfields}
 As mentioned, it is known that second-class constraints can be associated with gauge symmetry if one introduces new physical canonical degrees of freedom \cite{Henneaux:1992ig,Batalin:1986aq,Batalin:1986fm,Batalin:1991jm,Fradkin:1977xi,Fisch:1989ke}.
    In this approach, one modifies the second-class constraints by adding contributions from the new canonical degrees of freedom, so as to make the modified constraints first-class. We review this approach in detail in App.~\ref{sec:rhocanonical}. As we show there, second-class constraints can be traded for first-class gauge symmetry in a new theory, which reproduces the original configurations when the new degrees of freedom are constrained. This approach invokes new physical degrees of freedom, since the new canonical variables have observable effects, leading one in principle to predict a broader set of configurations that are not captured in the phase space of the original gauge theory. 

 As we discuss in App.~\ref{sec:rhocanonical}, in the case of shadow charges, one can introduce new canonical variables $\tilde{\rho}_a$ endowed with non-trivial Poisson brackets:
 \begin{align}\label{eq:rhocanon}
  \{\tilde{\rho}_a,\tilde{\rho}_b\}\equiv C_{abc}\tilde{\rho}_c\;.
 \end{align}
    (These Poisson brackets can be extended to products of $\tilde{\rho}_a$ by imposing the validity of Leibniz rule.)
 Furthermore, using $\{\tilde{\rho}_a,H\}=0$, one can replace the second-class constraints of Eq.~\eqref{eq:secondclassYM} with new first class constraints: \begin{align}\label{eq:firstclassYM}\tilde{\phi}_{1a}=G_a-\tilde{\rho}_a\approx 0\;,\quad\{\tilde{\phi}_{1a},\tilde{\phi}_{1b}\}=C_{abc}\tilde{\phi}_{1c}\approx 0\;.\end{align}
 These constraints will reproduce the configurations of Eq.~\eqref{eq:constrYM} when the initial conditions $\tilde{\rho}_a\approx\rho_a(\vec{x})$ are chosen.
 The $\tilde{\rho}_a$ are new, physical variables that correspond to the charge density of a charged field that is decoupled from the dynamics, meaning that its charge density is individually conserved, $\partial_t\tilde{\rho}_a=0$. For this to be the case, the new field that gives rise to $\tilde{\rho}_a$ must have infinite inertial mass. For these reasons, this approach corresponds to describing the shadow charges in terms of a UV completion of the theory, in which other fields are added at infinitely high mass. While this approach is feasible and usually regarded as a technical trick, it comes at the price of having to use extra fields to describe the full phase space of the gauge theory. In other words, in this approach, the presence of shadow charges appears to be a UV-dependent question. 
 Even more, one can see that theories with UV fields of high but finite mass cannot reproduce the exactly constant-in-time shadow charges predicted by gauge theory. In fact, a charged field with finite mass can always be made time-dependent by transferring momentum to it. Therefore, only a theory with truly infinitely heavy fields could reproduce the predictions of gauge theory. It is unclear whether the Hamiltonian of such a theory would be well defined, and whether such a degree of freedom could be described consistently when gravity is turned on.
    This shows that, despite the interpretation of shadow charges as background charge densities of heavy UV fields may be an intuitive step, it is non-trivial to argue that such theories exist. This point of view leaves the presence and consistency of shadow charges dependent on whether a theory of fields with strictly infinite inertial mass can be formulated.
 
 In contrast with this point of view, we now show that shadow charges and their symmetries can be described consistently without the need of introducing infinitely massive UV fields. In this new approach, gauge symmetry is left unbroken by the shadow charges and is generated by the second-class constraints, without changing any aspect of the theory. 
 
 This is the case because constraint surfaces connected by transformations generated through second-class constraints are physically equivalent. In other words, the second-class constraints only signal a gauge degeneracy at the level of the initial conditions.
 We present our argument in the next section, focusing on the case of the Yang-Mills shadow charges. As we will discuss, our considerations extend to any set of second-class constraints.
 
 \subsubsection{Shadow charges as purely IR data}\label{sec:constraintorbit}
    As we have discussed above, the fact that second class constraints change the constraint surface is usually regarded as a lack of symmetry of the configuration. In the approach we presented in Sec.~\ref{sec:UVfields} and App.~\ref{sec:rhocanonical}, the symmetry is retrieved in a system that closely resembles the gauge theory, but that contains new physical UV fields.
 
 In contrast with this view, we find it is actually possible to extract a symmetry from the second-class constraints of the system, without modifying the theory.
    This is akin to obtaining a purely IR description of the shadow charge, in which UV physics can be entirely ignored. This is possible if we simply recognize that not all of the initial conditions that define a specific constraint surface are observable. As it turns out, some of the initial conditions are physically indistinguishable. In fact, from the point of view of the Hamiltonian, constraint surfaces that are mapped into each other by the action of the second-class constraints are completely degenerate. This is because the Hamiltonian has vanishing Poisson brackets with the second-class constraints. Therefore, it is natural to view second-class constraints as gauge transformations that map physically equivalent initial conditions (\textit{i.e.}, constraint surfaces) into each other.
 In other words, second-class constraints only signal a degeneracy -- a gauge dependence --  in initial conditions, rather than a breaking of gauge symmetry.
 For instance, the constraint surface that we have defined in Eq.~\eqref{eq:constrYM} has information on the local orientation in color space of the adjoint vector $\rho_a(\vec{x})$. Due to the invariance of the Hamiltonian, configurations characterized by a different orientation are physically indistinguishable.
 
 With this in mind, rather than restricting our focus on the single constraint surface, we will find it useful to consider a broader object, which we call the constraint orbit, defined as the union of all of the physically equivalent constraint surfaces. This object will be invariant under the action of all of the $G_a$, meaning that it will capture the full gauge symmetry of the system. Following this understanding, in Sec.~\ref{sec:orbitBRST} we will introduce a generalization of the BRST symmetry that allows one to quantize second-class constraints without modifying the canonical structure of the theory.
 
 We now characterize the constraint orbit in the case of shadow charges in Yang-Mills theory.
 In order to describe this, we can note that there are residual gauge-invariant, first-class combinations of the constraints that correspond to the Casimir operators of the group. This happens because the matrix of Poisson brackets between the constraints, $C_{abc}\rho_c$ has zero eigenvalues. This degeneracy indicates residual gauge invariance of the constraint surface. For instance, the quadratic Casimir $G^2=G_a^2$ is a gauge invariant quantity:
 \begin{align}
  \{G^2,\phi_{1b}\}=2G_aC_{abc}G_c=0\;.
 \end{align}
 Correspondingly, one can find the following quadratic Casimir constraint:
 \begin{align}\label{eq:Cas2}
  {\Phi}_2= G_a^2-\rho_a^2\;,
 \end{align}
 which will be first class and generate the same transformations as the linear combination $2\rho_a\phi_{1a}$:
 \begin{align}
  \{\Phi_2,\phi_{1a}\}=0\;,\quad\text{and}\quad\{\Phi_2,F\}\approx2\rho_a\{\phi_a,F\}\;,
 \end{align}
 for a generic quantity $F$. 
 Depending on the group, the matrix $C_{abc}\rho_c$ will have multiple null eigenvectors. For instance, one could have a further independent first-class constraint corresponding to the cubic Casimir operator:
 \begin{align}\label{eq:Cas3}
  G^3=d_{abc}G_{a}G_{b}G_{c}\;,\quad{\Phi}_3= d_{abc}\,(G_a G_b G_c-\rho_a\,\rho_b\,\rho_c)\;,
 \end{align}
 with $d_{abc}=2\text{Tr}(\{T_a,T_b\}T_c)$ the totally symmetric tensor. Again, this cubic Casimir constraint generates the same transformations as the linear combination $3d_{abc}\rho_a\,\rho_b\,\phi_{1c}$:
 \begin{align}
  \{\Phi_3,\phi_{1a}\}=0\;,\quad\text{and}\quad \{\Phi_3,F\}\approx 3d_{abc}\rho_a\rho_b\{\phi_{1a},F\}\;.
 \end{align}
 Depending on the group, there may be non-trivial higher-order Casimir operators and corresponding first-class Casimir constraints. We will denote them $\Phi_J$, with $J$ indicating the order of the Casimir operator appearing in the constraint.
 
 The presence of these invariants is helpful in order to understand how to describe the constraint orbit. In fact, this is the orbit of a given constraint surface under the action of the $G_a$. Since both the constraint surface and its transformations are defined by the $G_a$, then it is clear that the constraint orbit will be completely characterized by the Casimir constraints, e.g.:
 \begin{align}\label{eq:constrOrb}
  \Phi_2\approx G_a^2-\rho_a^2\approx 0\;,&\quad\Phi_3\approx d_{abc}(G_aG_bG_c-\rho_a\rho_b\rho_c)\approx 0\,,
 \end{align}
 together with the temporal constraints $\phi_{0a}=\Pi_{a0}\approx0$. Higher-order Casimir constraints will have to be included, depending on the symmetry group. From this discussion, we see that the constraint orbit generalizes in a natural and consistent way the situation of first-class constraint surfaces. Indeed, the constraint orbit is a surface that shares the same invariance properties as the Hamiltonian, much like a constraint surface in the case of first-class constraints. On this surface, the non-dynamical quantities define orbits of physically equivalent field configurations, \textit{i.e.}, gauge symmetry. The only difference with respect to the case of first-class constraint surfaces is that now some of the initial conditions of the constrained fields are physically equivalent. For instance, in the case of Yang-Mills shadow charges, the initial conditions of the Casimir operators, Eq.~\eqref{eq:constrOrb}, carry all of the physical information in the initial conditions of all of the $G_a$.
 This discussion shows that the presence of shadow charges in Yang-Mills theory does not alter the gauge symmetry of the theory: as gauge transformations will always be generated by the $G_a$, regardless of the shadow charge.
 As we will show in Sec.~\ref{sec:gaugefixing}, this symmetry will be manifest at the level of the action.
    This result will be of relevance for the quantization of the theory, as having control on the gauge symmetry makes it possible to extract gauge-invariant information about the system and to discard unphysical divergences associated with pure gauge modes.
    Compared to the results of Sec.~\ref{sec:UVfields}, the present approach does not rest on imagining a UV theory of infinitely massive charges.
 
 Since we now know that shadow charges are consistent with gauge symmetry, we understand that the number of physical degrees of freedom is independent of the shadow charge.
 In fact, we can verify this by counting the number of dynamical and physical Hamilton equations for our system. Starting with the canonical variables $A_{ai}\,,\,\Pi_{ai}\,,\,A_{a0}\,,\,\Pi_{a0}$, we have $2\times4(N^2-1)$ Hamilton's equations for the $SU(N)$ theory. Among these, $2(N^2-1)$ are only apparently dynamical equations and can be traded for the non-dynamical constraints $\phi_{0a}\,,\,\phi_{1a}$. Moreover, since all of the constraints generate gauge transformations, $2(N^2-1)$ of the independent variables that transform under the constraints are not physical, and can therefore be discarded when computing observables. Therefore, one is left with $4(N^2-1)$ physical and dynamical Hamilton's equations, which implies that for a given shadow charge, the dynamical excitations are described by $4(N^2-1)$ initial conditions, corresponding to $N^2-1$ vector fields with two physical polarizations each. 
 
 The following construction is applicable in any situation in which second-class constraints appear, since the constraint orbit can be defined regardless of the algebra of the constraints. The main caveat in general will be that identifying the constraint orbit might not be as simple as in the case considered here. In general, one might need to explicitly integrate the gauge transformations of the constraint surface in order to find the constraint orbit.
	
\section{Describing the gauge-invariant dynamics}\label{sec:gauge}	
	So far, we have shown that the phase space of gauge theory contains configurations characterized by the presence of shadow charges. These shadow charges, we have argued, are fully compatible with gauge symmetry.
 Given these results, we now want to address two relevant technical aspects. One is extracting the physical information on the dynamics in a systematic way, imposing the constraints and removing gauge dependence from the observables.
 The other is showing how the gauge variables can be set to arbitrary values without affecting the physics, deriving a Lagrangian description of the dynamics in the presence of shadow charges.
    As we will discuss in Sec.~\ref{sec:quantum}, these results are useful for quantizing the theory in the presence of shadow charges, as one will need to discard gauge contributions in order to single out and regularize the observables of the theory.
 
 We address these points by constructing an object that carries information on both gauge symmetry transformations and shadow charges. In the case of usual gauge theory with vanishing shadow charges, this object is the BRST charge, which we will review below (see also Ref.~\cite{Barnich:2000zw,Becchi:1974md,Becchi:1974xu,Tyutin:1975qk,Becchi:1975nq}). This charge makes it simple to split the algebra of phase space in three sets of functions: gauge invariant, gauge dependent, and pure gauge functions. With this splitting, one can easily single out observables and gauge-fix the dynamics in a consistent way.
 As we will show in Sec.~\ref{sec:BRST}, the usual BRST construction will extend straightforwardly to the case of first-class constraints, e.g. electromagnetism with non-vanishing shadow charge and Yang-Mills shadow charges described in terms of UV fields. Instead, in order to obtain a consistent description of Yang-Mills shadow charges in purely IR terms, we will find that a natural generalization of the BRST charge is needed. We discuss this new construction in Sec.~\ref{sec:orbitBRST}. This generalization will account for the fact that gauge invariance is only captured by the constraint orbit, rather than the gauge dependent constraint surfaces.
 
 The BRST operator and its generalization will allow us to both describe the reduced phase space of observables in a simple way and to gauge-fix the dynamics in consistent ways. We will show this in Sec.~\ref{sec:gaugefixing}. In particular, the gauge-fixing will make it possible to introduce a non-trivial dependence of the Hamiltonian and the action on $A_0$ and the $A_{a0}$, showing how these variables can be introduced and discarded without changing physical predictions.
 
 In practical terms, once the BRST charge is given, $Q$, one can easily classify any function on phase space $F$ in terms of two simple operations:
 \begin{align}\label{eq:BRSTconds}\begin{split}
   1)&\;\;\text{is}\;\;F\;\;\text{BRST-invariant,}\quad\{F,Q\}=0\;?\\
   2)&\;\;\text{is}\;\;F\;\;\text{BRST-exact,}\quad F=\{V,Q\}\;\;\text{for some}\;V\;?
  \end{split}
 \end{align}
 The observables, \textit{i.e.}, the gauge-invariant quantities that are compatible with the constraints, will be those quantities that are BRST-invariant but not BRST-exact. {This set is usually described as a quotient of BRST-invariant quantities modulo the BRST-exact ones.} Quantities that are not BRST-invariant will be gauge dependent, while quantities that are BRST-exact will be pure gauge.
 
 This classification makes both gauge invariance and the arbitrariness of gauge redundancies explicit. Indeed, once the BRST charge is constructed, one can fix the dynamics of gauge variables in arbitrary ways by adding a non-observable, BRST-exact term $\{K,Q\}$ to the Hamiltonian. This is the gauge-fixing procedure:
 \begin{align}
  H\to H_K=H+\{K,Q\}\;.
 \end{align}
 This change in Hamiltonian does not affect observables, as the BRST-exact term induces an evolution that drops out of the quotient {that defines observable quantities}. Indeed, for a BRST-invariant quantity $A$, we find that the difference in time evolution between the original and the gauge-fixed Hamiltonian is BRST-exact:
 \begin{align}\label{eq:gaugefixinvariance}
  \{A,H_K\}-\{A,H\}=-\{K,\{Q,A\}\}-\{Q,\{A,K\}\}=\{V,Q\}\;,
 \end{align}
 where $V=\{K,A\}$.
 We will explore this construction and the consistent choices of $K$ in Sec.~\ref{sec:gaugefixing}.
 
 The reader can regard the construction of the BRST operator and its generalization, in Sec.~\ref{sec:BRST} and Sec.~\ref{sec:orbitBRST}, as a mathematical excursion that motivates introducing canonical ghost variables in elementary terms. This is especially useful in order to find the correct charge for the case of second-class constraints and their constraint orbit. {We will call this object the orbit-BRST charge.}
 The reader might want to skip those sections and only read about gauge-fixing. In that case, it can be useful to look up the notation for the ghost variables and their conjugate momenta, Eq.~\eqref{eq:ghostcanon}, and the expressions for the BRST and orbit-BRST charges, Eq.~\eqref{eq:abBRST} and \eqref{eq:orbitBRST}, with additional ghosts for the orbit-BRST defined in Eq.~\eqref{eq:casimirghosts}.

 \subsection{Reduced phase space and its resolution}\label{sec:BRST}
    In the following, we address the problem of extracting the physical information from a classical gauge theory. We do so by following the BRST construction in the classical phase space. This approach will be used through the rest of the work to gauge-fix and quantize the theory, and will be extended in a new direction in Sec.~\ref{sec:orbitBRST}, where we will treat the case of shadow charges in Yang-Mills. 

 Given a gauge theory described in terms of first-class constraints and respective gauge transformations, we want to reach a description of the system such that the constraints are solved and only gauge-invariant quantities are left. In such a description, there is no unphysical degeneracy associated with constraints and gauge transformations, and the only variables left are the physical observables.
 In principle, one could describe the reduced phase space of observables by finding a basis for all of the gauge-invariant quantities that can be built out of the canonical variables. However, finding such a complete set is quite difficult and may come at odds with manifest locality in space-time. 
 Moreover, one would need to restrict the theory to the constraint surface in order to avoid ambiguities in which a gauge invariant observable may be shifted by terms proportional to the constraints, with no observable difference. 
 
 Interestingly, rather than explicitly solving the constraints and finding appropriate gauge invariant variables, it is possible to define a simple operation defined on the whole phase space of the theory, whose result singles out the physical observables.
 This operation is the action of a differential $s$ (\textit{i.e.}, a nilpotent derivation, $s^2=0$), the so-called BRST differential, see e.g. \cite{Henneaux:1985kr,Barnich:2000zw,Becchi:1974md,Becchi:1974xu,Tyutin:1975qk,Becchi:1975nq,Fradkin:1975cq,Batalin:1977pb,Batalin:1981jr,Batalin:1983pz,Batalin:1985qj,Batalin:1989dm,Batalin:1983ggl,Anderson:1951ta}.    
 As we will see, it will be possible to represent the action of this differential canonically in terms of the so-called BRST charge $Q$:
 \begin{align}\label{eq:brstcanon}
  sF=\{F,Q\}\;,
 \end{align}
 where $F$ is any function of the canonical variables.
 Through this differential $s$, one can express the reduced phase space of observables as explained in Eq.~\eqref{eq:BRSTconds}, in terms of a quotient of quantities that are annihilated by the differential (BRST-closed), modulo quantities that are in the image of this differential (BRST-exact).
 (In technical terms, this quotient is called the cohomology of the BRST differential.)
 We outline here this construction, referring to e.g. \cite{Henneaux:1985kr,Barnich:2000zw} for a more detailed derivation.
 
 We will obtain the BRST differential $s$ as the combination of two distinct differentials, which respectively take care of quotienting out ambiguities related to terms proportional to the constraints and gauge redundancies. 
 This reflects the fact that two conditions single out the observables of the theory, as these are all the gauge invariant functions with support on the constraint surface: 
 \begin{align}\label{eq:cond}
  F\in C^\infty(\{\phi_\alpha\approx 0\})\;,\quad \text{such that}\quad \{F,\phi_\alpha\}\approx 0\;,
 \end{align}
 where $\phi_\alpha$ are all of the first-class constraints of the theory.
 The BRST construction is based on the fact that each of the two conditions in Eq.~\eqref{eq:cond} can be captured by the action of a differential, and that these two differentials can be combined into a new one. 
 As we will see, in order for these differentials to be defined, one needs to extend the phase space with auxiliary, unphysical variables, the so-called ghosts.
 Counting the powers of these additional variables, the so-called ghost number, makes it simple to recover the physical information from the extended phase space. In fact, physical information will be contained in quantities with zero ghost number. This counting provides what is called a grading of the algebra of functions in the extended phase space.
 
 To see how the first condition in Eq.~\eqref{eq:cond} is phrased in terms of a differential, we can notice that the algebra on functions with support on the constraint surface is given by the quotient algebra of functions over the whole phase space that differ by quantities that vanish on the constraint surface:
 \begin{align}\label{eq:reducedalgebra}
  C^\infty(\{\phi_\alpha\approx 0\})=\frac{C^\infty(P)}{\mathcal{N}(\{\phi_\alpha\approx 0\})}\;,
 \end{align}
 $\mathcal{N}(\{\phi_\alpha\approx 0\})$ being the set of functions vanishing on the constraint surface, and $P$ the original phase space of the theory.
 This quotient can be associated with quantities being invariant or exact under the action of a differential $\delta$.  Technically, a quantity $F$ invariant under this differential is said to be in its Kernel, $\text{Ker}(\delta)$, as it satisfies: $\delta F=0$. Instead, a quantity $F$ is exact with respect to $\delta$ if $F=\delta V$ for some quantity $V$. Then technically $F$ is said to be in the Image of $\delta$, $\text{Im}(\delta)$. 
    In terms of these notions, we want to associate the quotient of Eq.~\eqref{eq:reducedalgebra} to a quotient of the Kernel of $\delta$ by its image. If $\delta$ was only defined on the original phase space, then the requirement that its Kernel contains the algebra $C^\infty(P)$ would make $\delta$ trivial. Therefore, in order to make the image of $\delta$ equal to $\mathcal{N}(\{\phi_\alpha\approx0\})$, one needs to define $\delta$ as acting on an extended phase space containing some auxiliary variables $\mathcal{P}_\alpha$ associated with the constraints $\phi_\alpha$. For reasons that will become clear, these variables are called the ghost momenta and are counted with a negative ghost number.
    Once these variables are introduced, one can reproduce the space of functions on the constraint surface by taking the quotient of the zero ghost number part of the Kernel of $\delta$ to the zero ghost number part of its image:\footnote{(This quotient is called the zero ghost number homology of $\delta$.)}
 \begin{align}\label{eq:homology}
  \frac{\text{Ker}(\delta)_0}{\text{Im}(\delta)_0}=\frac{C^\infty(P)}{\mathcal{N}(\{\phi_\alpha\approx 0\})}\;,
 \end{align}
    where the subscript $0$ indicates taking the zero ghost number component of the Kernel and image.
 In practice, the action of $\delta$ on the extended phase space can be defined as:
 \begin{align}\label{eq:deltadef}
  \delta \mathcal{P}_\alpha=\phi_\alpha\;,\quad\delta|_P=0\;,
 \end{align}
    where $\delta|_P$ indicates the action of $\delta$ over the original phase space $P$. While this condition ensures that Eq.~\eqref{eq:homology} is satisfied, it is only when the $\mathcal{P}_\alpha$ are Grassmannian that $\delta$ becomes nilpotent, $\delta^2=0$.
    This can be understood simply in the case of a single constraint $\phi$, as non-trivial functions can be at most linear in the ghost momentum $\mathcal{P}$ if this is a Grassmannian variable, leading to $\delta^2=0$.
    We discuss how this works in more complex cases through an explicit example in App.~\ref{app:resolution}. There, we will see that when the Kernel of $\delta$ contains quantities with non-zero ghost number, these quantities are automatically $\delta$-exact, as a result of the Grassmannian nature of the ghost momenta.
    The construction of this differential $\delta$ is usually called the Koszul-Tate resolution or algebraic resolution of the constraint surface \cite{Henneaux:1992ig,Barnich:2000zw}.
 
 The differential $\delta$ allows one to discard non-observable ambiguities related to quantities that differ by terms proportional to the constraints. Once this is done, one can further enlarge the phase space and define a new differential that quotients out gauge transformations on the constraint surface. This is called the longitudinal differential, $d$, and directly parameterizes gauge transformations. 
 This differential can be represented in phase space by contracting the constraints with new canonical Grassmann variables, which we call ghosts and denote as $c_\alpha$. 
 With this extension of phase space, the longitudinal differential $d$ will act in phase space as:
 \begin{align}
  d F=c^\alpha\{F,\phi_\alpha\}\;.
 \end{align}
    As explained in \cite{Henneaux:1992ig,Barnich:2000zw}, the new auxiliary variables $c_\alpha$ represent a basis of the differential forms that can be associated with derivations along the gauge directions in phase space.
 Restricting $d$ to the constraint surface, one can single out quantities that are invariant along the gauge orbits by computing the quotient of the kernel of $d$ by its image:
 \begin{align}
  \{\text{gauge invariant observables on the constraint surface}\}=\frac{\text{Ker}(d)}{\text{Im}(d)}\;.
 \end{align}
 While $d$ can be extended to the whole phase space, it will in general be nilpotent only on the constraint surface: $d^2\approx 0$, but $d^2\neq 0$ in general.
 Therefore, $d$ is called a differential modulo $\delta$ over the whole phase space $P$.
 
 Having built the algebraic resolution of the constraint surface through $\delta$ and having defined an algebraic way to quotient gauge transformations through the longitudinal differential $d$, one can combine the two differentials into a new differential, defined on all phase space:
 \begin{align}\label{eq:brstdiff}
  s=\delta+d+\dots\;,
 \end{align}
 where the dots indicate additional terms of higher order in the $\mathcal{P}_\alpha$, which might be needed in order to ensure $s^2=0$. This is the BRST differential. 
 Once one has found this differential, the quotient of its Kernel (BRST-invariant quantities) by its image (BRST-exact quantities) at zero ghost number singles out the gauge invariant observables for a given constraint surface, e.g. for a given shadow charge.
 
 In order for $s$ to be a differential, $s^2=0$, one needs the longitudinal differential to satisfy compatibility conditions with the algebraic resolution given by $\delta$:
 \begin{align}
  \delta d+d\delta=0\;,\quad d^2=\delta s_1+s_1\delta\;,
 \end{align}
 with $s_1$ an operator linear in the $\mathcal{P}_\alpha$. The form of $s_1$ will then determine the higher order terms in Eq.~\eqref{eq:brstdiff}, see e.g. \cite{Henneaux:1992ig}.
 Importantly, the BRST differential $s$ can be represented canonically, as in Eq.~\eqref{eq:brstcanon}. For this to be the case, it is necessary to extend the Poisson brackets in such a way to make $c_\alpha$ and the $\mathcal{P}_\alpha$ conjugate to each other, assigning positive ghost number to the $c_\alpha$:
 \begin{align}\label{eq:ghostcanon}
  \{c^\alpha,\mathcal{P}_\beta\}=-\delta^\alpha_\beta\;.
 \end{align}
 Here, the Poisson brackets in the presence of conjugate bosonic variables $q\,,\,p$ and conjugate fermionic variables $\theta\,,\,\pi$ are defined as:
 \begin{align}
  \{A,B\}=\Bigg(\frac{\partial A}{\partial q}\frac{\partial B}{\partial p}-\frac{\partial A}{\partial p}\frac{\partial B}{\partial q}\Bigg)+(-)^{\varepsilon_A}\Bigg(\frac{\partial^L A}{\partial\theta}\frac{\partial^L B}{\partial\pi}+\frac{\partial^L A}{\partial\pi}\frac{\partial^L B}{\partial\theta}\Bigg)\;,
 \end{align}
 {with $\varepsilon_A=0$ or $1$ depending on whether $A$ is bosonic or fermionic, and the left derivative $\partial^L_\theta A$ defined by the variation of $A$ written with the variation of $\theta$ on the left, $A(\theta+\Delta\theta)\simeq A(\theta)+\Delta\theta\, \partial^L_\theta A$.}
 
 For instance, in an abelian gauge theory, e.g. when the Poisson brackets between the constraints vanish everywhere, $s$ is represented by the following BRST charge:
 \begin{align}\label{eq:abBRST}
  \quad Q=\int d^3 x\;c^\alpha\phi_\alpha\;,
 \end{align}
 which gives $s F=\{F,Q\}\;$.
 Indeed, we have that $s\mathcal{P}_\alpha=\{\mathcal{P}_\alpha,Q\}=\phi_\alpha$ and that $s F=dF$ on the original phase space $P$. 
 Instead, for a non-abelian theory with first-class constraints, in which e.g. $\{\phi_\alpha,\phi_\beta\}=B_{\alpha\beta\gamma}\phi_\gamma\approx 0$, one will have:
 \begin{align}\label{eq:nonabBRST}
  Q=\int d^3 x\;\Bigg(\,c^\alpha\phi_\alpha+\frac{1}{2}B_{\alpha\beta\gamma}\,c^\alpha\,c^\beta\mathcal{P}_\gamma\Bigg)\;.
 \end{align}
 This corresponds to setting $d\mathcal{P}_\alpha=B_{\alpha\beta\gamma}\,c^\gamma\mathcal{P}_\beta$, which is required in order to make $s$ nilpotent over the whole phase space, $\{Q,Q\}=0$. The extra term proportional to $\mathcal{P}_\alpha$ is an example of higher-order term of Eq.~\eqref{eq:brstdiff}. This is for example, the case for shadow charges in Yang-Mills theory described by introducing new canonical variables, as in Sec.~\ref{sec:UVfields}.
 With these results, one can verify that the BRST charge $Q$ is conserved:
 \begin{align}
  \frac{d}{dt}Q=\{Q,H\}= 0\;.
 \end{align}
 This fact reflects the gauge (and BRST) invariance of the Hamiltonian, \textit{i.e.}, its commutation with the constraints, as well as its independence on the ghost variables. 
 This conservation can also be seen as an indication of a global fermionic symmetry of the theory, the BRST symmetry.
 
 Practically, the BRST charge can be used to impose physical conditions on quantities dependent on the canonical variables (as well as on states, when the theory is quantized). In fact, having defined the BRST charge, we will have that the space of physical observables at a given shadow charge is given by all quantities that are BRST-invariant, up to BRST-exact terms:
 \begin{align}\label{eq:physcond}
  \{F,Q\}=0\;,\quad F\sim F+\{V,Q\}\;.
 \end{align}
 The construction of $Q$ can be carried out straightforwardly whenever the constraints are first class, \textit{i.e.}, whenever they have vanishing Poisson brackets with each other on the constraint surface:
 \begin{align}
  \{\phi_\alpha,\phi_\beta\}\approx 0\;.
 \end{align}
 This is the case for the constraint surfaces with non-vanishing shadow charge in electromagnetism. Instead, in the case of Yang-Mills theory, we will see that further considerations about the phase space are necessary in order to characterize the gauge invariant quantities at fixed shadow charges.
 
 The construction of the BRST differential makes it clear that the shadow charge density is a gauge-invariant observable, and gives a simple way to characterize the reduced phase space of observables in a gauge theory. As we will discuss in Sec.~\ref{sec:quantum}, the BRST construction is also helpful when quantizing the theory.
 
 Having outlined the construction of the BRS charge in the case of first-class constraints, in the next section, we extend this construction to the constraint orbit, \textit{i.e.}, the case in which gauge transformations are generated by second-class constraints.

 \subsection{The orbit-BRST}\label{sec:orbitBRST}
 As discussed in Sec.~\ref{sec:clYM}, Yang-Mills shadow charges define a constraint surface characterized by first and second-class constraints. While second-class constraints have been historically seen as a lack of gauge symmetry in the theory, as we have discussed, they have a natural interpretation as generators of gauge symmetry once one recognizes that the local orientation of the shadow charge in color space is not an observable.
    This corresponds to describing shadow charges consistently as purely IR features of theory, without the need to introduce new physical variables that are connected to non-standard UV physics.
 In the previous section, we have seen how the BRST differential can be defined in phase space to simultaneously impose gauge invariance as well as a chosen set of first-class constraints.
 
 Now, we want to construct a similar object for the case of shadow charges in Yang-Mills, in which second-class constraints make the constraint surface gauge dependent. As explained in Sec.~\ref{sec:constraintorbit}, gauge invariance can be retrieved by considering the set of all physically equivalent constraint surfaces, the constraint orbit. As we have discussed, the constraint orbit is gauge invariant and described by the Casimir constraints of Eq.~\eqref{eq:constrOrb}, together with the temporal constraints $\phi_{0a}=\Pi_{a0}\approx0$.
 
 While the single constraint surfaces are not invariant under the action of the $G_a$, the whole constraint orbit is mapped onto itself by these quantities.
 
 Already at the classical level, it is simple to recognize the reduced phase space of quantities defined on the constraint orbit and that are invariant under the action of the $G_a$. For instance, a subset of these are the functions that can be written as $\mathcal{F}=\delta(G_a^2-\rho_a^2)F(G_a^2)$, where $F(G_a^2)$ are functions with gauge invariant coefficients that depend on the canonical variables.
 Similarly to what we have done in the previous section, one can proceed by finding an algebraic resolution of this reduced phase space, \textit{i.e.}, a new differential, which we call the orbit-BRST, which 
 characterizes the reduced phase space of observables on the constraint orbit as a quotient of invariant quantities modulo exact ones. (In other words, the reduced phase space of observables on the constraint orbit will be given by the cohomology of the orbit-BRST differential.)
 This is the natural and physical extension of the usual BRST differential to situations with second-class constraints.
 
 To see more precisely how this construction is different from the standard BRST described in the previous section, note that while the $\Pi_{a0}$ lead to as many constraints as gauge transformations, the generators of spatial gauge transformations, the $G_a$, outnumber the Casimir constraints, which are the defining equations of the constraint orbit. Therefore, we have a departure from the usual case in which these two sets coincide. Nonetheless, we can still consistently go through the BRST approach and obtain a resolution of the constraint orbit. Following Sec.~\ref{sec:BRST}, we start by defining two differentials: the Koszul-Tate differential $\delta$, and the longitudinal differential $d$ on the constraint orbit, generated by the $G_a$ and $\Pi_{a0}$. As discussed in Sec.~\ref{sec:BRST}, the first differential discards the ambiguity corresponding to shifts of the observables by terms that vanish on the constraint orbit. Instead, the second differential generates translations in the unphysical gauge directions, which leave the constraint orbit invariant. Importantly, we will show that we can combine these two differentials and obtain a consistent quotient in terms of $d$ modulo $\delta$, encoding the reduced phase space of gauge invariant functions on the constraint orbit. This will be the orbit-BRST differential:
 \begin{align}
  s=\delta+d\;,
 \end{align}
 where no further terms are needed, since we will find that $d^2=0$.
 
 We now want to build the two differentials $\delta$ and $d$ explicitly in an extended phase space. This is especially important as it will allow us to construct an orbit-BRST charge as well as to confirm that the two differentials can be summed into a new differential, without the need to add terms of higher ghost number.
 
 The extended phase space will need to include ghost conjugate pairs for each of the generators of the two differentials. Unlike in the standard BRST, ghosts associated with $\delta$, \textit{i.e.}, the ones projecting quantities on the constraint orbit, will be unrelated to those associated with $d$, which instead parameterize gauge transformations.
 For the construction of $\delta$ we will start by introducing as many ghost momenta as there are Casimir constraints, as these will correspond to the defining equations of the constraint surface -- $\Phi_J\approx 0$. We will call these ghost momenta $\xi_J$. We will introduce the corresponding ghosts later. The action of $\delta$ is then defined by:
 \begin{align}
  \delta \,\xi_J=\Phi_J\;,
 \end{align}
 with $\delta$ acting trivially on the original phase space. 
 Similarly, we should introduce ghost momenta for the $\phi_{0a}=\Pi_{a0}$, which we call $\bar{c}_a$. We extend $\delta$ to these canonical variables by imposing:
 \begin{align}
  \delta \,\bar{c}_a=\Pi_{a0}\;.
 \end{align}
 Taking the ghost momenta to be Grassmannian, this definition makes $\delta$ a differential that singles out all the functions with support on the constraint orbit in terms of the zero ghost number quotient:
    \begin{align}
    \frac{\text{Ker}(\delta)_0}{\text{Im}(\delta)_0}=C^{\infty}(\{\Phi_J\approx 0\})\;.
    \end{align} 
 For the longitudinal differential $d$, in addition to the $\Pi_{a0}$ contributions, we will need to introduce $(N^2-1)$ ghost pairs, corresponding to all of the gauge transformations generated by the $G_a$. These are the same ghosts that one introduces for configurations with no shadow charge. We call the $G_a$--ghost pairs $c_a\,,\,\mathcal{P}_a$. Importantly, in this case, the $\mathcal{P}_a$ are not associated with the Koszul-Tate differential $\delta$, which only acts on the ghost momenta $\xi_J$ associated to the Casimir constraints. As for the temporal gauge transformations, we introduce the $\Pi_{a0}$--ghosts, which we call $\bar{\mathcal{P}}_a$:
 \begin{align}\label{eq:ghosts0}
  \{\bar{\mathcal{P}}_a,\bar{c}_b\}=\delta_{ab}\;.
 \end{align}
 Then, the longitudinal differential can be represented as:
 \begin{align}
  d \,F=\Bigg\{F,\int d^3x\Big(\bar{\mathcal{P}}_a\Pi_{a0}+c_aG_a+\frac{1}{2}C_{abc}c_ac_b\mathcal{P}_c\Big)\Bigg\}\;.
 \end{align}
 This definition grants that $d^2=0$ in all phase space. Since the $\Phi_J$ are abelian, we have that $\delta$ and $d$ are compatible: $\delta d+d\delta=0$. This ensures that we can sum the two differentials into a new one, the orbit-BRST differential $s=\delta+d$.
 
 At this point, it is very useful to note that the longitudinal differential can be modified by adding terms proportional to the Casimir constraints:
 \begin{align}\label{eq:dshift}
  dF\to dF+\Bigg\{F,\int d^3 x\;\chi_J\Phi_J\Bigg\}\;,
 \end{align}
 with $\chi_{J}$ any Grassmann function that has zero Poisson brackets with the $G_a\,,\,\Pi_{a0}\,,\,c_a\,,\,\mathcal{P}_a\,,\,\bar{\mathcal{P}}_a\,,\,\bar{c}_a$. This definition keeps $d^2=0$ and preserves the compatibility condition between $\delta$ and $d$: $\delta d+d\delta=0$.
 
 Given these differentials, we want to find a phase space representation of $s$ in terms of a canonical generator, the orbit-BRST charge.
 Finding an expression for this charge is not a priori an easy task, as representing $\delta$ on its own is not straightforward. 
 However, this is possible by further extending the phase space and adding ghost variables conjugate to the $\xi_{J}$. In fact, using the shift of the longitudinal differential Eq.~\eqref{eq:dshift} and setting $\chi_{J}$ as new canonical variables conjugate to the $\xi_{J}$,
 \begin{align}\label{eq:casimirghosts}\{\,\chi_{I},\xi_{J}\}=\delta_{IJ}\;,\end{align} 
 we obtain an orbit-BRST charge that represents the sum $s=\delta+d$ over the whole phase space:
 \begin{align}\label{eq:orbitBRST}
  Q=\int \!d^3 x\; \Bigg(\chi_{J}\Phi_J+\bar{\mathcal{P}}_a\Pi_{a0}+c_aG_a+\frac{1}{2}C_{abc}c_ac_b\mathcal{P}_c\Bigg)\;,\quad sF=\{F,Q\}\;.
 \end{align}
 Indeed, we have that
 \begin{align}
  \{\xi_{J},Q\}=\Phi_J=\delta \xi_{J}\;,\quad\text{and}\quad \{F,Q\}=dF\;,
 \end{align}
 for all quantities $F$ independent on the $\xi_{J}$. 
 
 In this construction, similarly to the standard case presented in Sec.~\ref{sec:BRST}, it is crucial to extend the phase space and introduce an algebraic structure between the ghost momenta $\xi_{J}$ and the newly introduced ghosts $\chi_{J}$ -- see e.g. Eq.~\eqref{eq:ghostcanon}. Note that this added structure in the extended phase space preserves the vanishing of graded brackets between $\delta$ and $d$, as we have:
 \begin{align}\label{eq:dshiftpres}
  \delta\{\,\chi_{I}\Phi_{I},\,\xi_{J}\}=\delta\Phi_J=0\;,\quad\text{and}\quad\{\,\chi_{I}\Phi_{I},\delta \xi_{J}\}=\{\,\chi_{I}\Phi_{I},\Phi_{J}\}=0\;.
 \end{align}
 The newly introduced Casimir ghost pairs have the effect of enforcing the Casimir constraints on the quotient defined by the orbit-BRST. This is different from the role of the remaining, usual ghost fields, which instead cancel gauge redundant modes. 
 
 With this, we have proved that the BRST construction can be extended to the case of non-vanishing shadow charges in a non-abelian theory, and more generally to systems with second-class constraints. This finding allows us to move beyond the gauge-dependent constraint surfaces that appear in these cases, and consider the reduced phase space associated to the whole constraint orbit. 
 Focusing on a properly invariant notion of constrained configuration space, the orbit-BRST represents a natural and physical extension of the BRST construction outside of the slice of phase space with vanishing shadow charges. As we will see, this realization is particularly useful when quantizing the theory, as it allows one to sidestep the problem of finding an operator representation of the algebra of Dirac brackets.
 As we discussed in Sec.~\ref{sec:constraintorbit}, this approach shows that shadow charges can be naturally treated as physical features of the gauge field configuration, without being associated with additional physical degrees of freedom. In other words, it is completely consistent to predict shadow charges regardless of the UV content of the theory.

    Before closing this section, we should stress the difference between the BRST differential at zero shadow charge and the orbit-BRST differential at non-zero shadow charge in a non-abelian theory. As we can see from Eq.~\eqref{eq:orbitBRST}, the difference between the orbit-BRST charge and the standard non-abelian BRST charge of Eq.~\eqref{eq:nonabBRST} is the Casimir term $\int d^3x\;\chi_J\Phi_J$. This term becomes trivial for $\rho_a=0$, as the Casimir operators generate trivial gauge transformations on the constraint surface with vanishing shadow charges, e.g.:
    \begin{align}
        \{F,G_a^2\}=2G_a\{F,G_a\}\approx 0\;,\quad\text{for}\quad G_a\approx0\;.
    \end{align}
    Instead, for non-vanishing shadow charges, the Casimir term changes the surface that is resolved by the quotient $\frac{\text{Ker}(s)_0}{\text{Im}(s)_0}$, setting this surface to be the constraint orbit with non-vanishing $\rho_a$. As we will see, this effect will be reflected on the physical conditions on the Hilbert space of the quantized theory.
    
 Having constructed the BRST charge and its generalization to the case of second-class constraints, we now move to showing how gauge theory with non-zero shadow charges can be gauge-fixed.

 \subsection{Gauge-fixed dynamics in the extended phase space}\label{sec:gaugefixing}
 So far, we have discussed the presence of shadow charges in the phase space of gauge theory and shown that these configurations are compatible with gauge symmetry. In the previous sections, we have further constructed the BRST charge $Q$ for abelian shadow charges and its generalization for the non-abelian case. With this construction, the constrained and gauge-invariant dynamics can be retrieved from an unconstrained system in the extended phase space of standard canonical variables plus ghosts and ghost momenta. This is by discarding all quantities that are not BRST invariant or that are BRST-exact.
 
 Using this setup, we now show how the dynamics of the gauge-redundant canonical variables can be changed arbitrarily without changing the dynamics of the gauge-invariant observables. 
    This is done by adding non-observable terms to the gauge-invariant (and BRST-invariant) Hamiltonian.  As discussed at the beginning of Sec.~\ref{sec:gaugefixing}, a simple way to make sure that the terms added are not observable is to choose them to be BRST-exact: $H\to H_K=H+\{K,Q\}$ for some $K$ appropriately chosen.
 The non-observable terms $\{K,Q\}$ will fix the dynamics of the gauge degrees of freedom, realizing a gauge fixing of the theory.
 
 Our choice of $K$ will be dictated by requiring the Hamiltonian to remain hermitian, bosonic and to have zero ghost number.
 This can be obtained by choosing $K$ to be a Grassmannian, anti-hermitian quantity with ghost number opposite to $Q$, known as the gauge-fixing fermion. In general, $K$ can be obtained by summing ghost momenta multiplied by quantities that depend on the variables of the original phase space:
 \begin{align}
  K=\int d^3 x\;f_\alpha(p,q)\mathcal{P}_\alpha\;,
 \end{align}
 where we indicate with $q$ and $p$ the original canonical variables.
 
 For a given gauge-fixed Hamiltonian $H_K$, one can immediately write down the corresponding gauge-fixed action $S_K$ by taking the Legendre transform:
 \begin{equation}\label{eq:gauge_fixed_action}
  S_K(q, \dot{q}, c, \dot{c}) = \int dt\, \left( \dot{q}\, p + \dot{c}\, \mathcal{P} - H - \{K, Q\} \right) \;.
 \end{equation}
 As we show below, this derivation allows to recast the theory into the more familiar form of a Faddeev–Popov action.
 Due to the dependence of the BRST charge $Q$ on the shadow charge, \textit{i.e.}, the configuration of the fields, the gauge fixed action and Hamiltonian will be configuration dependent, unlike the gauge and BRST invariant Hamiltonian that we have discussed so far.
 We now describe the gauge-fixing procedure in the cases of abelian and non-abelian shadow charges.

 \subsubsection{Gauge-fixing in electromagnetism}
 We now explicitly derive the gauge-fixed Hamiltonian and action for electromagnetism in the presence of a shadow charge. In doing so, we show how $A_0$ can be introduced or removed both in phase space and in the Hamiltonian without changing the physics. 
 
 We take as starting point the minimal Hamiltonian of Eq.~\eqref{eq:H}:
 $$
 H=\int d^3 x\,\Bigg(\frac{1}{2}\vec{\Pi}^2+\frac{1}{2}(\vec{\nabla}\!\wedge\!\vec{A})^2\Bigg)\;.
 $$
 
 As explained in Sec.~\ref{sec:BRST}, we can encode the choice of initial conditions of Eq.~\eqref{eq:constrrho} as well as the requirement of gauge invariance in the action of the BRST differential, by adding a ghost pair $(\,c, \mathcal{P})$ for the constraint $\phi_1$ and an antighost pair $(\,\bar{\mathcal{P}},\bar{c}\,)$ for the constraint $\phi_0$, with $\{ \,c(\vec{x}\,),\mathcal{P}(\vec{x}\,')\}=\{ \,\bar{\mathcal{P}}(\vec{x}\,),\bar{c}(\vec{x}\,')\} =  -\delta(\vec{x}-\vec{x}\,')$. Since the constraints are abelian, we have the following BRST charge operator:
 \begin{equation}\label{eq:BRSTnonmin}
  Q = \int d^3 x \, (-i \, \bar{\mathcal{P}} \, \phi_0 + c \, \phi_1) \,.
 \end{equation}
 The peculiar choice of ghost number chosen for $(\bar{c}\,,\, \bar{\mathcal{P}})$ makes it natural for $Q$ to have a well-defined mass dimension, while keeping the usual mass dimension for all canonical variables and their conjugates.
 This BRST charge quotients the extended phase space down to the reduced phase space of observables associated with the constraint surface of Eq.~\eqref{eq:constrrho}.
 
 By construction, the Hamiltonian $H$ has vanishing Poisson brackets with the BRST charge: $\{H,Q\}=0$, meaning that it is observable. As mentioned, shifting this Hamiltonian by a BRST exact quantity does not alter the spectrum of the observables, see Eq.~\eqref{eq:gaugefixinvariance}. The next step is therefore to fix the gauge and ghost variables by choosing a gauge-fixing fermion $K$. 
 A simple choice is the following:
 \begin{align}
  K=\int d^3x\;\mathcal{P} A_0\;.
 \end{align}
 With this choice, the Hamiltonian becomes:
 \begin{align}
  H_K = H + \{K, Q\}=\int d^3x\;\Bigg(\frac{1}{2}\vec{\Pi}^2+\frac{1}{2}(\vec{\nabla}\!\wedge\!\vec{A})^2-A_0(\vec{\nabla}\!\cdot\!\vec{\Pi}-\rho(\vec{x}))+i\mathcal{P}\bar{\mathcal{P}}\Bigg)\;,
 \end{align}
 which depends on both $A_0$ and on $\rho$. This Hamiltonian has only a kinetic term for the ghosts, meaning that these are automatically decoupled from the dynamics. To find the action, we compute:
 \begin{align}\label{eq:gauge_fixed_action_QED}
  S_K =& \int d^4 x \left(\dot{\vec{A}}\, \vec{\Pi}+\dot{A}_0\,\Pi_0 + \dot{c} \,\mathcal{P} + \dot{\bar{c}} \,\bar{\mathcal{P}} - \mathcal{H}_K\right) \;.
 \end{align}
 where $\mathcal{H}_K$ is the gauge-fixed Hamiltonian density, $H_K=\int d^3x\,\mathcal{H}_K$. While $\Pi_0\approx 0$ can be simply discarded, the conjugate momenta $\vec{\Pi}$ must be eliminated using Hamilton's equations for $\vec{A}$:
 \begin{align}\label{eq:elimPi}
  \dot{\vec{A}}=\{\vec{A},H_K\}&=\vec{\Pi}+\vec{\nabla}A_0\;,
 \end{align}
 leading to the usual $\vec{\Pi}=\dot{\vec{A}}-\vec{\nabla}A_0$. Note that the invariance of this equation under the gauge transformations generated by $\phi_0\,,\,\phi_1$ specifies what conditions the gauge functions $u_0$ and $u_1$ must satisfy. In the present case, we see that it must hold:
 \begin{align}
  \{\dot{\vec{A}},u_0\phi_0+u_1\phi_1\}=\vec{\nabla}\dot{u}_1&=\{\vec{\Pi}+\vec{\nabla}A_0,u_0\phi_0+u_1\phi_1\}=\vec{\nabla}u_0\;.
 \end{align}
 This holds only if $u_1=\lambda$ and $u_0=\dot{\lambda}$, leading to the usual gauge transformation $A_\mu\to A_\mu+\partial_\mu\lambda$.
 Using Eq.~\eqref{eq:elimPi}, and eliminating the ghost momenta as well, we obtain the gauge invariant, but configuration dependent action:
 \begin{align}
  S_K=\int d^4x\;\Bigg(-\frac{1}{4}F_{\mu\nu}F^{\mu\nu}+J^\mu_\S A_\mu+i\bar{c}\,\ddot{c}\Bigg)\;,
 \end{align}
 where we have defined $J_\S^\mu = (\rho(\vec{x}),0,0,0)$, and one can discard the ghost kinetic term, as these are decoupled (and in this gauge non-dynamical). Gauge invariance of this action is granted by the time-independence of $\rho(\vec{x})$.
 Besides deriving this simple action, the gauge-fixing procedure described above can be used to obtain more complex forms of the action. For instance, we can choose:
 \begin{equation}\label{eq:QED_gauge_fixing_fermion}
  K = \int d^3 x  \left( i \bar{c} (\vec{\nabla}\!\cdot\!\vec{A}) - \mathcal{P} A_0\right) \,,
 \end{equation}
 in order to enforce the Lorenz gauge, $\partial_\mu A^\mu = 0$, in the resulting gauge-fixed action.
 Indeed, we will have:
 \begin{align}\label{eq:Hgf}
  H_K = H + \int d^3 x \, \left((\vec{\nabla}\!\cdot\!\vec{A}) \Pi_0 - i \bar{c} \, \vec{\nabla}^2 c + i \mathcal{P} \bar{\mathcal{P}}+ \phi_1 \,A_0  \right)  \;,
 \end{align}
 Taking the Legendre transform and eliminating the canonical momenta $(\mathcal{P}, \bar{\mathcal{P}},\vec{\Pi})$, e.g. using Hamilton's equations, we obtain the following action:   
 \begin{align}\label{eq:gauge_fixed_action_QED_FP}
  S_K =  \int d^4 x \left( -\frac{1}{4} F_{\mu\nu}F^{\mu\nu} + i \bar{c} \,\Box c + (\partial_\mu A^\mu) b + J_\S^\mu\, A_\mu \right) \,,
 \end{align}
 where we have used the common notation $\Pi_0\equiv b$ and defined $J_\S^\mu = (\rho(\vec{x}),0,0,0)$. Varying this action with respect to $b$ imposes Lorentz gauge.
 As in the previous case, this action coincides with the usual gauge-fixed action for electrodynamics when one introduces Faddeev-Popov ghosts in the path integral formalism, with the addition of a background current $J_\S^\mu = (\rho(\vec{x}),0,0,0)$, which selects a rest-frame. As we have just shown, this result emerges naturally by considering the phase space of the theory and its Hamiltonian. Working from the Lagrangian perspective, one can obtain the same by adding a current $J^\mu_\S$ by hand as a static background of immobile charges. This was explored in \cite{Jacobson:2015mra}. In this sense, the Hamiltonian formalism seems to describe the shadow charges $\rho(\vec{x})$ more naturally than the Lagrangian formalism, showing how these are predictions of the theory without including physical charged background fields, and that these can be described with a configuration-independent object, the Hamiltonian of Eq.~\eqref{eq:H}.
 
 \subsubsection{Gauge-fixing in Yang-Mills theory} 
 
 We now examine the gauge-fixing procedure in the case of Yang-Mills theory with non-vanishing shadow charge.
 As we have discussed, this case can either be described by introducing new canonical variables representing physical UV charges or by following the orbit-BRST approach. Here we will illustrate both approaches.
 
 We can start by considering the Hamiltonian Eq.~\eqref{eq:HYM} and the constraint surface of Eq.~\eqref{eq:constrYM}.
    Then, we can introduce the canonical variables $\tilde{\rho}_a$ of Eq.~\eqref{eq:rhocanon} and describe the system in terms of the first-class constraints of Eq.~\eqref{eq:firstclassYM}, $\tilde{\phi}_{1a}=G_a-\tilde{\rho}_a\approx 0$ subject to the condition $\tilde{\rho}_a\approx\rho_a(\vec{x})$.
    Introducing ghost variables, this surface and its gauge symmetry are captured in the extended phase space by a BRST charge of the form of Eq.~\eqref{eq:nonabBRST}:
    \begin{align}\label{eq:BRSTrhocanon}
    Q=\int d^3 x\;\Bigg(\,\bar{\mathcal{P}}_a\Pi_{a0}+c_a\tilde{\phi}_{1a}+\frac{1}{2}C_{abc}\,c_a\,c_b\mathcal{P}_c\Bigg)\;.
    \end{align}
    Again, by construction, we have that the Hamiltonian is observable: $\{H,Q\}=0$.
 With this, we can gauge-fix the system. For instance, we can choose:
 \begin{equation}\label{eq:YM_gauge_fixing_fermion}
  K = \int d^3 x  \left( i \;\bar{c}_a (-\partial_k A_{ak}) - \mathcal{P}_a A_{a0}\right) \,,
 \end{equation}
 leading to the following gauge-fixed Hamiltonian:
 \begin{align}\label{eq:HK}
  H_K =& H + \{K, Q\}\\ =& H + \int d^3 x \, \left(A_{a0}(G_a-\tilde{\rho}_a) + i \mathcal{P}_a \bar{\mathcal{P}}_a + C_{abc} \mathcal{P}_a A_{b0} c_c+\partial_k A_{ak} \, b_a - i \bar{c}_a \, \nabla^2 c_a  \right)  \, \,\nonumber
 \end{align}
    where we are using the common notation $\Pi_{a0}=b_a$.
 Taking the Legendre transform, we obtain the gauge-fixed action: 
 \begin{equation}\label{eq:SK}
  S_K = \int d^4 x \left( -\frac{1}{4} F^a_{\mu\nu}F_a^{\mu\nu}+A_{a\mu}J^{\mu a}_\S+(\partial_\mu A^\mu)_a b_a + i \partial_\mu \bar{c}_a D^\mu c_a  \right) \, ,
 \end{equation}
 that is the usual Faddeev–Popov action in Lorenz gauge, plus an additional shadow charge current $J^{\mu a}_\S = ({\rho}_a(\vec{x}), 0,0,0)$.
 Note that if we remove the term $\bar{c}_a (-\partial_k A_{ak})$ in $K$, then we will retrieve the usual gauge transformations of the Lorentz-invariant action. In fact, if we choose $K=\int d^3 x\mathcal{P}_a A_{a0}$ we obtain:
    \begin{align}
        \{K,Q\}= \int d^3 x \, \left(A_{a0}(G_a-\tilde{\rho}_a) + i \mathcal{P}_a \bar{\mathcal{P}}_a + C_{abc} \mathcal{P}_a A_{b0} c_c  \right)\;,
    \end{align}
    and we find:
    \begin{align}\label{eq:basicallyphi2}
    \dot{G}_a=\{G_a,H_K\}=-C_{abc}A_{b0}G_c\;,\quad
        \dot{\tilde{\rho}}_a=\{\tilde{\rho}_a,H_K\}=-C_{abc}A_{b0}\tilde{\rho}_c\;,
    \end{align}
    meaning that $(D_0G)_a=(D_0\tilde{\rho})_a=0$. This condition implies that the ghost-free part of the gauge-fixed action is invariant under transformations $A_{a\mu}\to A_{a\mu}+(D_\mu\lambda)_a$. 
    Indeed, setting the ghosts to zero, the gauge-fixed action will be:
    \begin{align}\label{eq:simpleactionYM}
        S_K=\int d^4 x \left( -\frac{1}{4} F^a_{\mu\nu}F_a^{\mu\nu}+A_{a\mu}J^{\mu a}_\S \right) \, .
    \end{align}
    While $F_{\mu\nu}^aF^{\mu\nu}_a$ is straightforwardly invariant, the term $A_{a\mu}J^{\mu a}_\S$ is invariant, up to a total derivative, due to the condition $(D_0\tilde{\rho})_a=0$.
    As we explain in App.~\ref{app:gaugesymm}, this form of the gauge transformations can also be found by examining the gauge-fixed Hamilton's equation for $A_{ai}$.
    
    As we have explained in Sec.~\ref{sec:orbitBRST}, one can obtain the BRST classification of phase space for Yang-Mills shadow charges even without introducing the new physical variables $\tilde{\rho}_a$. This approach shows that shadow charges do not need to be thought of in terms of infinitely heavy charges, and rather are more naturally described as features of pure gauge theory, without any non-standard UV field.
    Following this route, one can constructs the gauge-fixed Hamiltonian by the same strategy described above, using the orbit-BRST charge of
    Eq.~\eqref{eq:orbitBRST}.
 
 Before concluding this section, we can observe a relevant symmetry of the action of Eq.~\eqref{eq:simpleactionYM} and of the equations of motion, beyond the gauge symmetry discussed above and in App.~\ref{app:gaugesymm}. That is, the action enjoys a symmetry under constant shifts of the vector field $A_{a\mu}$, regardless of the value of the fields.
    This can be seen as due to the conservation of the global $SU(N)$ Noether current:
    \begin{align}
        J^\mu_a=C_{abc}A_{b\nu}F^{\mu\nu}_c\;,\quad \partial_\mu J^\mu_a=0\;,
    \end{align}
    where the second equation holds only on the equations of motion.
    If we perform a shift of the vector fields $A_{a\mu}$ by constant vectors $\ell_{a\mu}$, the action will change by:
    \begin{align}
    \delta_\ell S_K=\int d^4x \;\ell_{a\mu}(C_{abc}A_{b\nu}F^{\mu\nu}_c+&\rho_a(\vec{x})\delta^\mu_0)=\int d^4x\;\ell_{a\mu}(J_a^\mu+\delta^\mu_0\rho_a(\vec{x}))=\int d^4x\;\partial_\mu U^\mu\;,\\
    \text{with:}\quad U^\mu=&\;\ell_{a\nu}\,x^\nu (J_a^\mu+\delta^\mu_0\rho_a(\vec{x}))\;,
    \end{align}
    where we have used that $\partial_\mu(J_a^\mu+\delta^\mu_0\rho_a(\vec{x}))=0$. When the integral of $\partial_\mu U^\mu$ converges, this result implies that the equations of motion do not change under constant shifts of the vector fields $A_{a\mu}$. This means that zero-frequency modes solve the equations of motion even in the presence of background fields.

 Having discussed the classical aspects of shadow charges in gauge theory, we now turn to their quantization.

	\section{Quantization of shadow sectors}\label{sec:quantum}
     In this section, we show how the results obtained so far, at the classical level, can be translated to the quantum theory, leading to a consistent description of quantized shadow charges.

    The quantum theory is obtained by canonical quantization of the classical theory, in which one finds an operator representation of canonical variables such that graded Poisson brackets (indicated by curly brackets) are mapped into commutators or anticommutators (both indicated by square brackets): $\{\cdot\,,\,\cdot\}\to[\cdot\,,\,\cdot]$. The operators will act on a Hilbert space of quantum states of the system, which parameterize the configurations described in phase space. For simplicity of notation, we will indicate the field operators with the same symbols used for the corresponding canonical variables. Unless explicitly stated, in the following we will always work in terms of operators of the quantized theory, rather than classical fields.

    In the following, we first describe in  Sec.~\ref{sec:physcond} how to single out gauge-invariant observables in the quantum theory by means of a physical condition imposed on the canonical Hilbert space. This condition will single out a physical Hilbert space, over which quantum observables can be computed in a gauge-invariant way. In Sec.~\ref{sec:QED}, we will consider QED with a shadow charge, discussing the physical excitations of the quantized theory as well as the physical states and the path integral. In Sec.~\ref{sec:QYM} we will discuss shadow charges in the quantized non-abelian theory, following the two approaches presented in Sec.~\ref{sec:UVfields} and Sec.~\ref{sec:constraintorbit}, and demonstrating their physical equivalence. In particular, we will discuss expectation values over the physical states, as well as the path integral and the spectrum of the observables. This will clarify that Yang-Mills shadow charges must satisfy certain quantization conditions.

    \subsection{Physical condition on the Hilbert space}\label{sec:physcond}
    
    As usual for quantized gauge theories, it is desirable to discard gauge redundancies. This makes it possible to eliminate divergences due to exchange/counting of unphysical gauge modes, and allows us to sidestep the problem of finding a complete set of gauge invariant quantities. Discarding the gauge symmetry can be done by either or both gauge-fixing, \textit{i.e.}, imposing conditions that fix gauge variables by explicitly breaking gauge symmetry, and/or by imposing a physical condition on the states of the Hilbert space. In this work, we will avoid relying entirely on the gauge fixing, and rather impose a physical condition on the Hilbert space. Similarly to our discussion in the previous section, the physical condition will be encoded by the action of the BRST charge operator on the states, and will correspond to extracting the information of the reduced phase space of the observables.
    
    We now discuss the choice of physical condition. This condition should single out states that transform trivially under gauge transformations, in such a way that the physical Hilbert space is left with no dependence on the gauge variables. When the generators of gauge transformations coincide with the constraints, the physical condition may also select the initial conditions chosen for the constrained variables.
    For instance, one can impose gauge invariance of the physical states $|\psi\rangle$ by requiring that they change by an overall phase $e^{i\lambda_\alpha\phi_\alpha}|\psi\rangle=e^{i\lambda_\alpha\theta_\alpha}|\psi\rangle$, with $\theta$ a real number. This would imply: \begin{align}\label{eq:physcondnofock}\phi_\alpha|\psi\rangle=\theta_\alpha\ket{\psi}\;,\end{align} 
    as a physical condition. 
    This is a consistent choice for systems with first-class constraints, \textit{i.e.}, when the commutator of the constraints is vanishing or proportional to the constraints: $[\phi_\alpha,\phi_\beta]|\psi\rangle =B_{\alpha\beta\gamma}\phi_\gamma|\psi\rangle $. When this is the case, one can satisfy Eq.~\eqref{eq:physcondnofock} by setting e.g. $\theta_\alpha=0$ if $B_{\alpha\beta\gamma}\neq 0$, or with any choice of $\theta_\alpha$ if $B_{\alpha\beta\gamma}=0$.
    Instead, for systems with second-class constraints, Eq.~\eqref{eq:physcondnofock} is not always consistent.
    For instance, in the case of two second-class constraints of the form $\phi_q=q\approx 0$ and $\phi_p=p\approx 0$, with $[q,p]=1$, we would have:
    \begin{align}
        0=(\theta_q\theta_p-\theta_p\theta_q)\ket{\psi}=[\phi_q,\phi_p]\ket{\psi}=[q,p]\ket{\psi}=\ket{\psi}\neq0\;,
    \end{align}
    which is absurd.
    However, in the case of shadow charges in Yang-Mills theory, with second-class constraints given in Eq.~\eqref{eq:secondclassYM}, if one imposes $\phi_{1a}\ket{\psi}=(G_a-\rho_a)\ket{\psi}=\theta_a\ket{\psi}$, then it follows:
    \begin{align}\label{eq:contradict}
    0=(\theta_a\theta_b-\theta_b\theta_a)\ket{\psi}=[\phi_{1a},\phi_{1b}]|\psi\rangle=[G_a-\rho_a,G_b-\rho_b]|\psi\rangle= C_{abc}(\theta_c+\rho_c)|\psi\rangle\;,
    \end{align}
    which can hold only if $\theta_a=-\rho_a$, \textit{i.e.}, if $G_a\ket{\psi}=0$. In these computations, we are using that the $\rho_a(\vec{x})$, being initial conditions with trivial Poisson brackets, are mapped to operators proportional to the identity. If we impose Eq.~\eqref{eq:contradict}, we only obtain configurations in which the Casimir operators have vanishing expectation value, e.g.:
    \begin{align}
        \bra{\psi}G_a^2\ket{\psi}=0\;,
    \end{align}
    which fails to represent the configurations that we have found in the classical phase space.

    This makes it clear that Eq.~\eqref{eq:physcondnofock} cannot be used to consistently describe second-class constraints and non-abelian shadow charges. 
    As discussed around Eq.~\eqref{eq:diracbrack}, this problem could be in principle addressed by using Dirac brackets, in which the second-class constraints are mapped to operators proportional to the identity. However, as we will show in Sec.~\ref{sec:QYM}, it is practically unfeasible to find an operator representation of the classical Dirac's brackets algebra, which makes this approach impractical.

    In analogy with our classical discussion, two alternatives appear viable. One is to alter the canonical structure of the theory, introducing new physical variables as in Eq.~\eqref{eq:rhocanon} and App.~\ref{sec:rhocanonical}. In this approach, we substitute the initial conditions $\rho_a$, proportional to the identity, with non-trivial operators $\tilde{\rho}_a$ that have commutators
    $
        [\tilde{\rho}_a,\tilde{\rho}_b]=C_{abc}\tilde{\rho}_c\;.
    $
    As we explained, this corresponds to explicitly including decoupled charged fields with infinite inertial mass in the UV of the theory. When the constraints are modified to be $\tilde{\phi}_{1a}=G_a-\tilde{\rho}_a$, then we see that they become first class, and Eq.~\eqref{eq:physcondnofock} can be imposed consistently, e.g. $\tilde{\phi}_{1a}|\psi\rangle=0$. As we discussed in Sec.~\ref{sec:clYM}, the UV theory of these fields may not be well defined.
    
    The second way to impose the physical conditions on the Hilbert space follows our proposal of Sec.~\ref{sec:constraintorbit}, and avoids adding new physical operators in the theory. In practice, in this approach, we leave the initial conditions $\rho_a$ to be c-number functions, but we recognize that we should impose initial conditions in a gauge-invariant way. In this way,  the whole gauge symmetry hidden in the second-class constraints becomes manifest.
    Still, one cannot easily impose initial conditions as well as gauge invariance on the physical states. For instance, asking that the Casimir constraints $\Phi_J$ given e.g. in Eq.~\eqref{eq:constrOrb} satisfy Eq.~\eqref{eq:physcondnofock} is incompatible with asking their gauge invariance. Indeed, as explained above, one would have to impose $G_a\ket{\psi}=0$, leading to $G_a^2\ket{\psi}=0$, incompatible with the initial conditions. Of course, one could avoid this by not imposing gauge-invariance conditions with respect to all of the $G_a$, but this would result in a residual gauge symmetry in the Hilbert space.
    Given these difficulties, to understand what the physical condition on states should be, we find it useful to connect to our classical understanding of Sec.~\ref{sec:gauge}. There, we showed that the BRST charge $Q$ allows to single out the algebra of the observables as BRST-invariant quantities modulo BRST exact ones, see Eq.~\eqref{eq:BRSTconds}. When we quantize canonically the system, this quotient algebra becomes an algebra of operators that act on the canonical Hilbert space of the theory. {In the quantum theory, the classical BRST charge $Q$ becomes a nilpotent, hermitian operator, see e.g.~\cite{Kugo:1977yx,Kugo:1977zq,Kugo:1979gm}}.
    From this point of view, we understand that the physical Hilbert space should be the space over which the quotient algebra of the observables acts in a well-defined way. In other words, the physical Hilbert space should be such that BRST-exact operators act trivially and have vanishing matrix elements. 
    To show how these requirements are satisfied in practice, consider for instance an observable operator $E$. This is defined up to BRST exact terms, $E+[V,Q]$, {where $V$ is a generic operator}. Here the square brackets indicate the graded commutator of operators, \textit{i.e.}, the anticommutator in the case of fermionic operators.
    To start, we can require the matrix elements of any observable $E$ on physical states to be uniquely defined regardless of BRST-exact terms:
    \begin{align}
    \bra{\psi'}(E+[V,Q])\ket{\psi}=\bra{\psi'}E\ket{\psi}+\bra{\psi'}[V,Q]\ket{\psi}=\bra{\psi'}E\ket{\psi}\;,
    \end{align}
    for all the physical states $\bra{\psi'}\,,\,\ket{\psi}$ and for all operators $V$.
    This means that BRST-exact operators have vanishing matrix elements on physical states, and it can only be the case if all the states of the physical Hilbert space are annihilated by $Q$:
    \begin{align}\label{eq:condQ}
        Q\ket{\psi}=0\;.
    \end{align}
    Indeed one then has: $\bra{\psi'}[V,Q]\ket{\psi}=\bra{\psi'}(QV+VQ)\ket{\psi}=0$. States satisfying Eq.~\eqref{eq:condQ} are called BRST-closed.
    Not all the states in the canonical Hilbert space satisfy Eq.~\eqref{eq:condQ}, and there will be states $\ket{\varphi}$ such that $Q\ket{\varphi}=\ket{\chi}\neq 0$. States of the form $Q\ket{\varphi}$ are called BRST-exact. Such states are annihilated by the BRST operator, due to its nilpotence, $Q^2\ket{\varphi}=0$, meaning that they satisfy Eq.~\eqref{eq:condQ} and have zero norm:
    \begin{align}
        \bra{\varphi}Q \;Q\ket{\varphi}=0\;.
    \end{align}
    All these zero norm BRST-exact states have zero scalar product with all states satisfying Eq.~\eqref{eq:condQ}: $\langle\psi|Q|\varphi\rangle=0$. 
    This means that BRST-exact states are decoupled from the rest of the BRST-closed states satisfying Eq.~\eqref{eq:condQ}. As we will see, this decoupling is useful in order to characterize the theory, as it will imply that ghost and gauge excitations decouple from the physical ones, see Sec.~\ref{sec:quartet}.
    Given this decoupling, it is consistent and desirable to remove the BRST-exact states and define the physical Hilbert space to have a non-degenerate scalar product. This is obtained by identifying all the BRST-exact states with zero, meaning that we will identify states that differ by a BRST-exact state:
    \begin{align}\label{eq:BRSTquotientH}
        \ket{\psi}+Q\ket{\varphi}\sim\ket{\psi}\;.
    \end{align}
    Here, the tilde symbol $\sim$ indicates that two states are identified in the definition of the physical Hilbert space. 
    Bringing Eq.~\eqref{eq:condQ} and Eq.~\eqref{eq:BRSTquotientH} together, we obtain the following definition of physical states:
    \begin{align}\label{eq:BRSTconditionQ}
        Q\ket{\psi}=0\;,\quad \ket{\psi}+Q\ket{\varphi}\sim\ket{\psi}\;.
    \end{align}
    This definition corresponds to taking the physical states as the quotient of the Kernel of the BRST operator by its image in Hilbert space. This reflects the quotient definition of the reduced phase space of the observables, as we have seen in Sec.~\ref{sec:gauge}. Imposing the physical condition allows us to formulate the quantized gauge theory consistently, just by following canonical quantization. In this way, we sidestep the problems related to defining explicitly gauge-invariant states as in Eq.~\eqref{eq:physcondnofock}.
    Note that the definition of physical states in Eq.~\eqref{eq:BRSTconditionQ} implies that the action of observable operators on the physical Hilbert space is defined uniquely regardless of BRST-exact operators:
    \begin{align}
        (E+[V,Q])\ket{\psi}=E\ket{\psi}+QV\ket{\psi}\sim E\ket{\psi}\;.
    \end{align}
    In summary, the BRST approach relaxes the requirement of gauge invariance of the states, but produces a consistent operator representation of the algebra of the observables, with a well defined action on the physical Hilbert space.
    We will adopt the BRST point of view in the quantization of gauge theories with shadow charges, both in the abelian and non-abelian case. 
    
    As we discussed in Sec.~\ref{sec:gaugefixing}, the operator $Q$ will also allow us to gauge-fix the Hamiltonian and the action of the theory, making it possible for us to write the path integral in a convenient way.

 \subsection{QED}\label{sec:QED}
 We now canonically quantize the variables of the phase space of electrodynamics, $\vec{A}(\vec{x},t)\,,\,\vec{\Pi}(\vec{x},t)$ and $A_0(\vec{x},t)\,,$ $\Pi_0(\vec{x},t)$ by finding operators with same commutation relations as their Poisson bracket:
    \begin{align}
        [A_i(\vec{x},t),\Pi_j(\vec{y},t)]=\delta^3(\vec{x}-\vec{y})\;,\quad [A_0(\vec{x},t),\Pi_0(\vec{y},t)]=\delta^3(\vec{x}-\vec{y})\;.
    \end{align}
 Given these operators, the time-evolution operator will be given by the Hamiltonian operator written following Eq.~\eqref{eq:H}, reproducing the configurations studied in the previous sections. The shadow charges $\rho(\vec{x})$ will now be trivial operators proportional to the identity. As explained in Sec.~\ref{sec:BRST}, we can extend the Hilbert space and the set of operators by introducing ghosts, in such a way as to conveniently describe gauge invariant quantities in the presence of shadow charge, using the BRST charge operator of Eq.~\eqref{eq:abBRST}. In the following, Sec.~\ref{sec:quartet}, we will characterize the physical states in the presence of a shadow charge in terms of their excitations. We will show that the physical conditions of Eq.~\eqref{eq:BRSTconditionQ} imply the decoupling of unphysical gauge and ghost modes. This reflects the fact that the canonical variables corresponding to these unphysical modes are not part of the reduced phase space of the observables, which is selected by the BRST quotient. In the quantum theory, the gauge and ghost modes associated with each pair of constraints combine in such a way as to decouple in quartets.
    We will then discuss the path integral in Sec.~\ref{sec:pathintQED}.
    
 \subsubsection{The quartet mechanism for shadow charges}\label{sec:quartet}
 Here, we show how the quantum theory reflects the fact that the BRST quotient captures the physical information on the system. In particular, in the quantum theory, the physical Hilbert space can be shown to have zero number of unphysical ghost and gauge excitations. This decoupling of the unphysical modes, by quartets of two gauge and two ghost modes, is called the quartet mechanism and has been explored in the standard cases of vanishing shadow charges in e.g. \cite{Kugo:1979gm,Kugo:1977yx,Kugo:1977zq,Nakanishi:1996he}.
 In order to show this, we want to define an operator $N$, with eigenvalues $n$, counting the number of unphysical excitations, gauge redundancies plus ghost excitations. It will turn out that this operator is BRST exact: $N=\{K,Q\}$, for a certain operator $K$. This has the relevant implication that the physical Hilbert space of BRST-invariant states quotiented by BRST-exact states, defined in Eq.~\eqref{eq:BRSTconditionQ}, coincides with the eigenstates of $N$ with $n=0$ eigenvalue, \textit{i.e.}, states with zero unphysical excitations.
 
 In order to show this, we can start by constructing the creation and annihilation operators corresponding to the fields in play. We define an invertible decomposition of the bosonic fields $(A_\mu, \Pi^\mu)$ in terms of the following creation/annihilation operators $a_0, a_{T1}, a_{T2}, a_L, a_0^\dagger, a_{T1}^\dagger,a_{T2}^\dagger, a_L^\dagger$ as follows:
\begin{align}\label{eq:bosonicdecomposition}
A_0(\vec{x},t)=&\int \frac{d^3\vec{k}}{(2\pi)^{3/2}}\frac{1}{\sqrt{2\omega_k}}\Big(a_0(\vec{k},t)e^{i\vec{k}\cdot\vec{x}}+a_0^\dagger(\vec{k},t)e^{-i\vec{k}\cdot\vec{x}}\Big)\;,\nonumber\\
   \Pi_0(\vec{x},t)=&\int \frac{d^3\vec{k}}{(2\pi)^{3/2}}{i}\sqrt{\frac{\omega_k}{2}}\Big[(a_0(\vec{k},t)+a_L(\vec{k},t))e^{i\vec{k}\cdot\vec{x}}-(a_0^\dagger(\vec{k},t)+a_L^\dagger(\vec{k},t))e^{-i\vec{k}\cdot\vec{x}}\Big]\;,\nonumber\\
  \vec{A}(\vec{x},t)=&\int \frac{d^3\vec{k}}{(2\pi)^{3/2}}\frac{1}{\sqrt{2\omega_k}}\Big(\sum_{i=1,2}\vec{\epsilon}_{Ti}(\vec{k})a_{Ti}(\vec{k},t)e^{i\vec{k}\cdot\vec{x}}+\frac{\vec{k}}{|\vec{k}|}a_L(\vec{k},t)e^{i\vec{k}\cdot\vec{x}}+h.c.\Big)\;,\nonumber\\
  \vec{\Pi}(\vec{x},t)=&\int \frac{d^3\vec{k}}{(2\pi)^{3/2}}i\sqrt{\frac{\omega_k}{2}}\Big(\sum_{i=1,2}\vec{\epsilon}_{Ti}(\vec{k})a_{Ti}(\vec{k},t)e^{i\vec{k}\cdot\vec{x}}+\frac{\vec{k}}{|\vec{k}|}(a_0(\vec{k},t) + a_L(\vec{k},t))e^{i\vec{k}\cdot\vec{x}}-h.c.\Big) + \vec{\Pi}_\rho(\vec{x})\;,
 \end{align}
    where $\vec{\epsilon}_{Ti}(\vec{k})$ are unit vectors orthogonal to $\vec{k}$, and $\vec{\Pi}_\rho(\vec{x})$ denotes a classical initial condition for the electric field, which will be proportional to the identity and which we will define more precisely later. In these expressions $h.c.$ denotes the hermitian conjugate. The decomposition presented provides a solution to the classical equations of motion of the theory, according to the gauge-fixed Hamiltonian of Eq.~\eqref{eq:Hgf}.
From the equal time commutation relation $[A_0(\vec{x},t),\Pi_0(\vec{y},t)]=\delta^3(\vec{x}-\vec{y})$, $[A_i(\vec{x},t),\Pi_j(\vec{y},t)]=\delta_{ij}\delta^3(\vec{x}-\vec{y})$,  we find that:
 \begin{align}
[a_0(\vec{k},t),a_0^\dagger(\vec{k}',t)]&=i\delta^3(\vec{k}-\vec{k}')\;\\
[a_L(\vec{k},t),a_L^\dagger(\vec{k},t)]&=i \delta^3(\vec{k}-\vec{k}')\;,\\
[a_{Ti}(\vec{k,t}),a^\dagger_{Tj}(\vec{k}',t)]&=i\delta_{ij}\delta^3(\vec{k}-\vec{k}')\;.
\end{align}
We then decompose the ghost fields in terms of creation/annihilation operators $g, g^\dagger, \bar{g}, \bar{g}^\dagger$ as follows:
\begin{align}
  c(\vec{x},t) = -\int \frac{d^3\vec{k}}{(2\pi)^{3/2}} \frac{1}{2\omega_k^{3/2}} (g \,e^{i\vec{k}\cdot\vec{x}} + g^\dagger e^{-i\vec{k}\cdot\vec{x}}) \,, \quad 
  \mathcal{P}(\vec{x},t)=i\int \frac{d^3\vec{k}}{(2\pi)^{3/2}} \omega_k^{3/2} (\bar{g} \,e^{i\vec{k}\cdot\vec{x}} + \bar{g}^\dagger e^{-i\vec{k}\cdot\vec{x}})\; , \nonumber\\
        \bar{c}(\vec{x},t) = -i\int \frac{d^3\vec{k}}{(2\pi)^{3/2}} \omega_k^{1/2} (\bar{g} \,e^{i\vec{k}\cdot\vec{x}} - \bar{g}^\dagger e^{-i\vec{k}\cdot\vec{x}}) \,, \quad 
  \bar{\mathcal{P}}(\vec{x},t)=-\int \frac{d^3\vec{k}}{(2\pi)^{3/2}} \frac{1}{2\omega_k^{1/2}} (g \,e^{i\vec{k}\cdot\vec{x}} - g^\dagger e^{-i\vec{k}\cdot\vec{x}})\; , 
 \end{align}
    where the factors $\omega_k$ have been chosen so as to keep the desirable mass dimensions for both the BRST charge $Q$ and the creation/annihilation operators. The choice of $g,g^\dagger, \bar{g}, \bar{g}^\dagger$ mixes the ghosts related to $\phi_0$ and those related to $\phi_1$, which is convenient for the purposes of this section.
We impose the equal-time anticommutation relations $[c(\vec{x},t),\mathcal{P}(\vec{y},t)]=[\bar{\mathcal{P}}(\vec{x},t),\bar{c}(\vec{y},t)]=-\delta^3(\vec{x}-\vec{y})$, where again we are using the square brackets to indicate a graded commutator, \textit{i.e.}, the anticommutator for fermionic operators. With this, we obtain:
\begin{align}\label{eq:commghost}
[g(\vec{k},t),\bar{g}^\dagger(\vec{k}',t)]&=[\bar{g}(\vec{k},t),g^\dagger(\vec{k},t)]=-i\delta^3(\vec{k}-\vec{k}') \,.
\end{align}
It is convenient to introduce the Fourier decomposition of the shadow charge
\begin{equation}
    \rho(\vec{x}) = \int \frac{d^3\vec{k}}{(2\pi)^{3/2}}\rho(\vec{k}) e^{-i\vec{k}\cdot\vec{x}} \, ,
\end{equation}
where $\rho(\vec{k})$ is proportional to the identity. The classical initial condition for the electric field $\vec{\Pi}_\rho(\vec{x})$, is chosen such that the dynamics satisfy the constraint $\vec{\nabla}\cdot \vec{\Pi} - \rho(\vec{x}) \approx 0$. Thus,
\begin{equation}\label{eq:initialvecPi0}
    \vec{\Pi}_\rho(\vec{x}) = \int \frac{d^3\vec{k}}{(2\pi)^{3/2}}\frac{\vec{k}}{|\vec{k}|}\frac{i\rho(\vec{k})}{|\vec{k}|} e^{-i\vec{k}\cdot\vec{x}} \,.
\end{equation}
The decomposition in Eq.~\eqref{eq:bosonicdecomposition}, with the choice in Eq.~\eqref{eq:initialvecPi0} enables a separation of the longitudinal background electric field from the dynamical wave modes of the theory. Consequently, the dispersion relation takes the simple form $\omega_k = |\vec{k}|$.

We now compute the BRST charge $Q = \int d^3 x \left(-i \bar{\mathcal{P}} \Pi_0 + c (\vec{\nabla}\cdot \vec{\Pi}-\rho(\vec{x})\right)$ substituting the field decompositions derived above and obtaining:
\begin{align}
  Q=\int d^3k\Bigg(g^\dagger(\vec{k},t)a(\vec{k},t)+g(\vec{k},t)a^\dagger(\vec{k},t)\Bigg)\;,
 \end{align}
    where we have introduced: 
\begin{align}
a(\vec{k},t)=a_L(\vec{k},t)+a_0(\vec{k},t)\;,\quad b(\vec{k},t)=\frac{1}{2}(a_L(\vec{k},t)-a_0(\vec{k},t))\;.
\end{align}
These operators satisfy the following commutation relations:
\begin{align}\label{eq:commut}
[a(\vec{k},t),b^\dagger(\vec{k}',t)]=i\delta^3(\vec{k}-\vec{k}')\;,\quad[b(\vec{k},t),a^\dagger(\vec{k}',t)]=i\delta^3(\vec{k}-\vec{k}')\;.
\end{align}
The decoupling of the unphysical modes is granted by an interplay between the quartet of unphysical modes
 $a_0\,,\,a_L\,,\,g\,,\,\bar{g}$. In the specific, mixing these quartets allows us to show that the operator $N$ counting the number of unphysical excitations is BRST exact. Indeed,  
    we can find an operator $K$ such that $N=[K,Q]$. To see this, we first express $N$ in terms of the operators $a\,,\,b\,,\,g\,,\,\bar{g}$ and their conjugates. Given the commutation relations of Eq.~\eqref{eq:commghost} and~\eqref{eq:commut}, it will be:
    \begin{align}
        N=\int d^3k\Big(a^\dagger b +b ^\dagger a+g^\dagger\bar{g}+\bar{g}^\dagger\,g\Big)\;.
    \end{align}
    Then, choosing the following operator,
    \begin{align}
        K=\int d^3k \Big(\bar{g}^\dagger(\vec{k},t) \,b(\vec{k},t)\,+b^\dagger(\vec{k},t)\,\,\bar{g}(\vec{k},t)\Big)\;,
    \end{align}
    we obtain:
 \begin{align}
  [K,Q]=&\Big[\int d^3k \Big(\bar{g}^\dagger(\vec{k},t) \,b(\vec{k},t)\,+b^\dagger(\vec{k},t)\,\,\bar{g}(\vec{k},t)\Big),Q\Big]\\
  &=\int d^3k\Big([\bar{g}^\dagger \,\,b,a^\dagger g_k]+[b^\dagger\,\,\bar{g},g^\dagger \,a]\Big)\\
  &=\int d^3k\Big(a^\dagger b +b ^\dagger a+g^\dagger\bar{g}+\bar{g}^\dagger\,g\Big)= N\;.
 \end{align}
 The operator $N$ counts the number of excitations of the unphysical fields, and as such its spectrum is given by non-negative integers.
 By nilpotence of $Q$ and the Jacobi identity, the fact that $N$ is BRST exact also implies $[N,Q]=0$. This means that the two operators are simultaneously diagonalizable, e.g. the Kernel of $Q$ will be expressible in terms of eigenspaces of $N$ with eigenvalues $n$. 
 Due to $N$ being BRST exact, we have that the BRST-invariant eigenstates of $N$ with $n\neq 0$, $|\psi_n\rangle$, are necessarily BRST exact:
 \begin{align}\label{eq:N=0}
  |\psi_n\rangle=\frac{N}{n}|\psi_n\rangle=\frac{1}{n}[K,Q]|\psi_n\rangle=Q\,\frac{K}{n}|\psi_n\rangle\equiv Q|\chi\rangle\;.
 \end{align}
 Therefore, all of the states with $n\neq0$ are projected out of the quotient, which is then formed solely by states with a vanishing number of unphysical components.
 
 This shows that the physicality condition $Q|\psi\rangle=0$, together with the quotienting by BRST exact states, grants the decoupling of unphysical modes from the dynamics. This is only made possible by completing the gauge pairs conjugate to the constraints on $\Pi_0$ and $\vec{\nabla}\cdot\vec{\Pi}$ to quartets including pairs of ghosts, which is why this is usually called the quartet mechanism.

 \subsubsection{Shadow charges and coherent states}\label{sec:coherent}
 In this section, we characterize the physical Hilbert space in the presence of shadow charges in terms of coherent states. We focus on the treatable case of shadow charges in QED, and we only comment on the case of Yang-Mills, where the explicit construction of physical states becomes more complex. Without loss of generality, we begin with the BRST operator defined in the phase space of $\vec{A}\,,\,\vec{\Pi}$:
 \begin{equation}
  Q(\rho) = \int d^3 x \, c \left(  \vec{\nabla} \cdot \vec{\Pi} - \rho(\vec{x}) \right) \,.
 \end{equation}
 The BRST-invariant Hamiltonian $H$ defined in Eq.~\eqref{eq:H} allows us to define the vacuum $\ket{0}$ as the unique state with zero energy and positive norm. This is the usual Lorentz-invariant vacuum state of gauge theories. Accordingly, this state is in the quotient identified by $Q (\rho=0)$. However, the Lorentz-invariant vacuum $\ket{0}$ is not annihilated by $Q (\rho)$, as 
 \begin{equation}
  Q (\rho) \ket{0} = Q(0) \ket{0} - \int d^3 x \, c \, \rho(\vec{x}) \ket{0}= - \int d^3 x \, c \, \rho(\vec{x})\ket{0}\neq 0 \,.
 \end{equation}
    This is non-vanishing because, by virtue of the quartet mechanism in the absence of shadow charge, the vacuum state $\ket{0}$ has zero ghost excitations, meaning that $c\ket{0}\neq 0$.
 We will show that the states annihilated by $Q (\rho)$ are instead given in the form of a coherent state of longitudinal electric field, excited on top of the Lorentz-invariant vacuum $\ket{0}$. A coherent state of a longitudinal electric field ($E^{\rm cl}_i  = \nabla_i \Phi^{\rm cl}$) can be constructed as~\cite{Berezhiani:2021zst} 
 \begin{align}
  \ket{\rho} = \exp\left\{i\int d^3 x \, E_j^{\rm cl} {A}_j\right\}\ket{0} \equiv e^{-i f} \ket{0}\, ,
 \end{align}
 where $E_j^{\rm cl}$ is a c-number function specifying the field configuration in the state $\ket{\rho}$, while $A_j$ represents, in our notations, the component of the field operator $\vec{A}$. In other words, $\bra{\rho}{\Pi}_j\ket{\rho} = E_j^{\rm cl}$. 
 Let us now act on this state with $Q (\rho)$,
 \begin{equation}\label{eq:step_1}
  Q (\rho) \ket{\rho} =  e^{-i {f}}  Q (\rho)  \ket{0} + \left[Q(\rho), -i {f}\right] e^{-i {f}} \ket{0} \, ,
 \end{equation}
 where we used the identity
 \begin{equation}
  [g({A}), {B}] = [{A}, {B}] \frac{\partial g(x)}{\partial x}\Big|_{x=A} \, ,
 \end{equation}
    for any operators $A\,,\,B$ such that $[A, [B,A]] = 0$. Since $\left[Q(\rho), -i {f}\right] = \int d^3 x \, c \, \vec{\nabla} \cdot \vec{E}_{\rm cl}$, Eq.~\eqref{eq:step_1} becomes:
 \begin{equation}
  Q (\rho) \ket{\rho} =  \int d^3 x \, c \left( -\rho +  \vec{\nabla} \!\cdot\! \vec{E}_{\rm cl} \right) \ket{\rho} \,.
 \end{equation}
 Thus, $\ket{\rho}$ is part of the physical Hilbert space identified by $Q (\rho)$ if $\vec{\nabla} \cdot\vec{E}_{\rm cl} = \rho$. Notice that, by construction, the state $\ket{\rho}$ has positive norm, since $\braket{\rho}{\rho} = \braket{0}{0} > 0$.
 
 This demonstrates that, in the quantum theory, configurations with a shadow charge can be described by coherent states of longitudinal electric field.
    Clearly, in addition to the coherent longitudinal electric field, allowed states can have transverse electromagnetic fields. For instance, one will have a coherent transversal component of the electric field, e.g. if $E_j^{\rm{cl}}$ is not purely longitudinal.
 Similarly, one can have coherent magnetic fields in states of the form:
 \begin{equation}
  \ket{J} = \exp\left\{-i\int d^3 x \left(A_j^{\rm cl} \Pi_j - E_j^{\rm cl} {A}_j\right)\right\}\ket{0} \, ,
 \end{equation}
 in which also $\vec{A}$ gets a non-trivial expectation value corresponding to a magnetic field. It is straightforward to verify that $Q (\rho) \ket{J} = 0$, provided again that $\vec{\nabla} \cdot\vec{E}_{\rm cl} = \rho$. Coherent gauge modes will instead be discarded by the BRST quotient, as they will be BRST exact, see Eq.~\eqref{eq:N=0}.
 
    This discussion suggests that coherent states might also describe physical states in the case of Yang-Mills shadow charges. While this seems likely, the non-linearity in the Gauss' law operators $G_a$ makes it non-trivial to build a coherent state $\ket{\rho}$ satisfying $Q\ket{\rho}=0$, see e.g.~\cite{Berezhiani:2021zst}. This problem may be solved by finding gauge-invariant dressing of gluon fields, which is however a difficult task.

 \subsubsection{Path integral in QED}\label{sec:pathintQED}
 Here we use the construction of the gauge-fixed action of Sec.~\ref{sec:gaugefixing} to write the path integral in the extended BRST phase space. This has the benefit that gauge invariance is replaced by a global BRST symmetry, making expectation values manifestly symmetric under changes of gauge-fixing conditions. Here we proceed in the opposite logical order compared to the traditional Faddeev–Popov method, where ghost fields are introduced to represent the Faddeev–Popov determinant, and the BRST symmetry emerges only as a byproduct of invariance of the functional measure of integration under gauge-fixing. In contrast with that view, here BRST symmetry comes before quantization and explicitly corresponds to discarding non-observable quantities from the theory.
 
 Starting from the gauge-fixed action Eq.~\eqref{eq:gauge_fixed_action_QED}, the generating functional of the theory reads:
 \begin{equation}
  Z_K = \int \mathcal{D} A_\mu  \, \mathcal{D} \Pi^\mu  \,\mathcal{D} c  \,  \mathcal{D} \bar{c} \, \mathcal{D} \mathcal{P} \, \mathcal{D} \bar{\mathcal{P}} \exp\Bigg(i \int d^4 x \left(\dot{A}_\mu \Pi^\mu + \dot{c} \,\mathcal{P} + \dot{\bar{c}} \,\bar{\mathcal{P}} - \mathcal{H}_K\right)\Bigg) \,,
 \end{equation}
    where in this expression and in the rest of this section, the variables indicate classical fields rather than quantum operators.
 We now perform the integral over $\mathcal{P}$ using the identity $\int \mathcal{D}  \mathcal{P} \exp\{i \int d^4 x \, \mathcal{P} (\dot{\,c} - i \bar{\mathcal{P}})\} \propto \delta(\bar{\mathcal{P}} - (-i \dot{c}))$. The resulting functional delta function simplifies the subsequent integration over $\bar{\mathcal{P}}$, yielding:
 \begin{align}\begin{split}
  Z = \int \mathcal{D} A_\mu  \, \mathcal{D} \Pi^\mu  \, \mathcal{D} c  \,  \mathcal{D} \bar{c} \,\text{exp}\Bigg(&i \int d^4 x \left( \dot{A}_0 \Pi^0 + \dot{A}_i \Pi^i + \dot{\bar{c}} (-i  \dot{c} )\right) \\
  &- \frac{1}{2} \Pi^i \Pi_i - \frac{1}{4} F_{ij}F^{ij}- \partial_k A^k b + i \bar{c} \,\nabla^2 c - \phi_1 A_0 \Bigg)\,.
 \end{split}\end{align}
 We further perform the integration over $\Pi^i$, which can be carried out exactly since the action is quadratic in $\Pi^i$. Defining $F_{0\,i} = \dot{A}_i -\partial_i A_0$, we finally obtain:
 \begin{align}
  Z = \int \mathcal{D} A_\mu  \, \mathcal{D} b  \, \mathcal{D} c  \,  \mathcal{D} \bar{c} \,\,\text{exp}\Bigg(i \int d^4 x \left( -\frac{1}{4} F_{\mu\nu}F^{\mu\nu} + i \bar{c} \,\Box c + (\partial_\mu A^\mu) b + \rho_\S(\vec{x})\, A_0 \right)  \Bigg) \,,
 \end{align}
 and thus we recover Eq.~\eqref{eq:gauge_fixed_action_QED_FP}. The integration over $b$ yields a $\delta(\partial_\mu A^\mu)$, \textit{i.e.}, restricts the integration over configurations obeying the Lorenz gauge. Considering more general choices than the gauge-fixing fermion of Eq.~\eqref{eq:QED_gauge_fixing_fermion}, allows one to recover the action in the more general $R_\xi$ gauge, see e.g.~\cite{Henneaux:1992ig}.

 \subsection{Quantized Yang Mills theory}\label{sec:QYM}
    In this section, we discuss aspects of the quantized Yang-Mills theory in the presence of shadow charges. In particular, in Sec.~\ref{sec:observables} we consider how the shadow charges affect the expectation values of the Gauss' law operators on physical states. In Sec.~\ref{sec:pathintegralYM}, we discuss the path integral of the theory, while in Sec.~\ref{sec:quantizedrho} we discuss the allowed values for the shadow charge density in the quantized theory.
    
    Before discussing these points, we briefly address the difficulty encountered if one attempts to eliminate the constraints by using Dirac brackets. As discussed around Eq.~\eqref{eq:diracbrack}, these brackets demote the constraint functions (or equivalently, the Gauss' law operators) to operators proportional to the identity, which have vanishing commutators with all the operators in the theory. This would imply that the gauge transformations become trivial, and that the gauge variables do not evolve.
 The Dirac brackets of the canonical variables, in the presence of shadow charges $\rho_a$ are:
 \begin{align}
  \{A,B\}_D=\{A,B\}-\{A,\phi_{1a}\}\mathcal{B}^{-1}_{ab}\{\phi_{1b},B\}
 \end{align}
 where $\mathcal{B}^{-1}_{ab}$ is the inverse of the matrix $C_{abc}\rho_c$ in a maximal non-degenerate eigenspace. For instance, for $SU(2)$, we can take $\rho_a=(\rho_1,0,0)$ and find:
 \begin{align}
  \mathcal{B}^{-1}_{ab}=\frac{i}{2}\begin{pmatrix}
   0 & 0 & 0\\
   0 & 0 & \frac{1}{\rho_1}\\
   0 & -\frac{1}{\rho_1} & 0
  \end{pmatrix}\;.
 \end{align}
 With this choice, we can compute explicitly the Dirac brackets of the canonical variables. We have:
 \begin{align}
  \{A_{a0}(\vec{x}),\Pi_{b0}(\vec{y})\}_D=&\delta_{ab}\delta^3(\vec{x}-\vec{y})\;,\\
  \{A_{ai}(\vec{x}),\Pi_{bj}(\vec{y})\}_D=&\delta_{ab}\delta_{ij}\delta^3(\vec{x}-\vec{y})\;,
 \end{align}
 In addition to these brackets, we see that the Dirac brackets between the different $A_{ai}$ are non-vanishing:
 \begin{align}
  \{A_{ai}(\vec{x}),A_{bj}(\vec{y})\}_D=\delta^3(\vec{x}-\vec{y})\Bigg(&C_{aec}A_{ei}(\vec{x})\mathcal{B}^{-1}_{cb}(\vec{x})\frac{\partial}{\partial y^j}-C_{bec}A_{ej}(\vec{x})\mathcal{B}^{-1}_{ca}(\vec{x})\frac{\partial}{\partial y^i}\;\Bigg.\\\Bigg.&+\mathcal{B}^{-1}_{ab}(\vec{x})\frac{\partial^2}{\partial y^i\partial y^j}+\frac{\partial \mathcal{B}^{-1}_{ab}(\vec{x})}{\partial x^j}\frac{\partial}{\partial y^i}+C_{aec}\mathcal{B}^{-1}_{cd}(\vec{x})C_{dfb}A_{ei}(\vec{x})A_{fj}(\vec{x})\Bigg)\;.\nonumber
 \end{align}
 The complex form of this operator makes it difficult to find an operatorial representation of the Dirac brackets.
    
    Given this difficulty, we find it useful to implement BRST quantization following either the approach of Sec.~\ref{sec:constraintorbit} or that of Sec.~\ref{sec:UVfields}, that is, imposing the physical conditions of Eq.~\eqref{eq:BRSTconditionQ}:
    $$Q|\psi\rangle=0\;,\quad|\psi\rangle+Q|\chi\rangle\sim|\psi\rangle\;,$$
    with $Q$ the operator given by either Eq.~\eqref{eq:orbitBRST}
    or Eq.~\eqref{eq:BRSTrhocanon}.
    With this construction, one can study time evolution and matrix elements among physical states using the gauge fixed Hamiltonian of e.g. Eq.~\eqref{eq:HK}. Since $[H_K,Q]=0$, time evolution maps the physical Hilbert space into itself. 

    \subsubsection{Physical states for non-abelian shadow charges}\label{sec:observables}
    We now want to characterize the physical states for Yang-Mills theory in the presence of shadow charges.
    As we have discussed, we can single out gauge-invariant observables in the case of Yang-Mills shadow charges in two ways. One is introducing new quantum operators corresponding to the charge density of UV fields with infinite inertial mass, as discussed in Sec.~\ref{sec:UVfields}, and then building a BRST charge as in Eq.~\eqref{eq:BRSTrhocanon}. In this case, the physical condition is:
    \begin{align}\label{eq:Qzerorho}
        \int d^3x \Bigg(c_a (G_a-\tilde{\rho}_a)+\frac{1}{2}C_{abc}c_a c_b\mathcal{P}_c+\bar{\mathcal{P}}_a\Pi_{a0}{+u_ab_a}\Bigg)\ket{\psi}=0\;.
    \end{align}
    Once this condition is imposed, the gauge-dependent $\tilde{\rho}_a$ can be gauge-fixed to the initial conditions $\rho_a$, as we will show in Sec.~\ref{sec:pathintegralYM}. {Here, compared to Eq.~\eqref{eq:BRSTrhocanon}, we have added ghost fields $u_a$ (with conjugate momenta $w_a$) and auxiliary fields $b_a$, which will be useful for the gauge-fixing in the next section. Similarly to $A_{a0}$, $\Pi_{a0}$, and the respective ghosts, these extra fields can be introduced in the BRST construction without {altering the physical content of the theory}.}
    As we mentioned, the $\tilde{\rho}_a$ should be interpreted as the charge density of a UV field with infinite inertial mass.
    The other approach is to impose only gauge invariant initial conditions on the system, as discussed in Sec.~\ref{sec:constraintorbit}, and to construct an orbit-BRST charge as in Sec.~\ref{sec:orbitBRST} and Eq.~\eqref{eq:orbitBRST}. In this case, the physical condition becomes:
    \begin{align}\label{eq:Qzeroorbit}
        \int d^3x\Bigg(\chi_J\Phi_J+c_a G_a+\frac{1}{2}C_{abc}c_a c_b\mathcal{P}_c+\bar{\mathcal{P}}_a\Pi_{a0}\Bigg)\ket{\psi}=0\;,
    \end{align}
    with $\Phi_J$ the Casimir constraints, e.g., for $SU(3)$:
    \begin{align}
    \Phi_2=G_a^2-\rho^2(\vec{x})\;,\quad\Phi_3=d_{abc}G_a G_b G_c-\rho_{(3)}^3(\vec{x})\;,
    \end{align}
    where $\rho^2\,$ and $\,\rho^3_{(3)}$ are indipendent functions and $d_{abc}=2\text{Tr}(\{T_a,T_b\}T_c)$ the totally symmetric tensor.
    Both approaches, at the classical level, allow one to extract the algebra of gauge invariant quantities compatible with the initial conditions imposed on the system. Naturally, the initial conditions selected will be reflected in the expectation values of the quantized theory, once the physical conditions of Eq.~\eqref{eq:Qzerorho} and Eq.~\eqref{eq:Qzeroorbit} are imposed.
    As we will examine, this leads to an apparent tension between the two approaches, \textit{i.e.}, between the two possible choices of BRST charge $Q$, as they lead to different expectation values for the Gauss' law operators $G_a$. This tension is due to the different initial conditions selected in the two approaches. The approach of Sec.~\ref{sec:UVfields} imposes gauge-dependent initial conditions, $G_a\approx\rho_a$, while the approach of Sec.~\ref{sec:constraintorbit} imposes only gauge-invariant initial conditions, e.g. $G^2\approx\rho^2$, as well as conditions on higher order Casimir combinations. 
    As we will see, gauge dependent conditions of the first approach can be seen as a gauge-fixing of the gauge-invariant initial conditions defined in the second approach.

    To proceed with this discussion, note that all BRST-exact operators will have vanishing expectation value on all physical states, as imposed by Eq.~\eqref{eq:BRSTconditionQ}. It is useful to examine the consequences of this fact in the two approaches presented.
    In the first case, in which we adopt the BRST operator of Eq.~\eqref{eq:BRSTrhocanon}, we will have:
    \begin{align}
        G_a-\tilde{\rho}_a+C_{abc} c_b\mathcal{P}_c=[\mathcal{P}_a,Q]\;.
    \end{align}
    This means that, considering a state $\ket{\psi}$ over which $C_{abc} c_b\mathcal{P}_c$ has vanishing expectation value (we will discuss alternatives below), we must have:
    \begin{align}\label{eq:physcondUV}
        \bra{\psi}(G_a-\tilde{\rho}_a)\ket{\psi}=0\;.
    \end{align}
    Then, choosing a gauge-fixing such that the expectation value of $\tilde{\rho}_a$ is equal to $\rho_a$, we will have $\langle G_a\rangle =\rho_a$, when the state has unit norm. On one side, this expectation value reflects the defining equations of the constraint surface at a non-zero shadow charge. On the other hand, it prescribes a non-zero expectation value to the gauge-dependent Gauss' law operators.
    We can contrast this with the constraint orbit approach, in which we only impose gauge-invariant Casimir constraints of Eq.~\eqref{eq:constrOrb}. In that case, the BRST charge is given by Eq.~\eqref{eq:orbitBRST}, and we have the following exact operators:
    \begin{align}
        G_a^2-\rho^2(\vec{x})=[\xi_2,Q]\;,\quad G_a+C_{abc} c_b\mathcal{P}_c=[\mathcal{P}_a,Q]\;,
    \end{align}
    where $\xi_2$ is the quadratic Casimir ghost, see Eq.~\eqref{eq:casimirghosts}. Again, considering physical states $\ket{\psi}$ such that $\langle C_{abc} c_b\mathcal{P}_c\rangle =0$, we then find the following expectation values:
    \begin{align}\label{eq:physcondorbit}
        \bra{\psi}(G_a^2-\rho^2(\vec{x}))\ket{\psi}=0\;,\quad\bra{\psi}G_a\ket{\psi}=0\;.
    \end{align}
    This seems a surprising finding, as one would expect physical states to allow for the description of classical states in which $\langle G_a\rangle\neq 0$. Such states certainly exist, but are not part of the physical Hilbert space defined in the second approach. Even more, one might worry that the two conditions in Eq.~\eqref{eq:physcondorbit} are incompatible with each other. That would be the case if the Gauss' law operators $G_a$ were themselves BRST-exact: $G_a=[V,Q]$, as one would obtain $\langle G_a^2\rangle=0$ on physical states, due to $Q^2=0$ and to Eq.~\eqref{eq:Qzeroorbit}. This however, cannot be the case, since we have $[G_a,Q]=C_{abc}c_bG_c\neq 0$. Indeed, if we had $G_a=[V,Q]$, it would be instead $[G_a, Q] = [[V,Q],Q]=0$, due to $Q^2=0$ and the Jacobi identity.
    
    Importantly, there will be states $\ket{\psi}$ that satisfy Eq.~\eqref{eq:physcondorbit}. A simple example of such state in $SU(2)$ is an eigenstate of $G_a^2$ and e.g. $G_{a=1}$ with eigenvalues $\rho^2(\vec{x})$ and $0$ respectively:
    \begin{align}
        G_a^2\ket{\rho,0}=\rho^2(\vec{x})\ket{\rho,0}\;,\quad G_{a=1}\ket{\rho,0}=0\;.
    \end{align}
    Since the $G_{a\neq 1}$ will map this state into states orthogonal to it, we will have zero expectation value for all $G_{a\neq 1}$. Taking the state to have unit norm, we have:
    \begin{align}
        \bra{\rho,0}G_a^2\ket{\rho,0}=\rho^2(\vec{x})\;,\quad \bra{\rho,0}G_a\ket{\rho,0}=0\;.
    \end{align}
    For the general case of $SU(N)$, we can similarly diagonalize simultaneously the Casimir operators and the combinations of $G_a$ that commute with each other, {\it i.e.}, the Cartan subalgebra. Assigning generic eigenvalues to the Casimir operators and zero eigenvalues to the generators in the Cartan subalgebra, we obtain again Eq.~\eqref{eq:physcondorbit}.
    With these results, it is clear that Eq.~\eqref{eq:physcondorbit} is self-consistent. 
    We can interpret this equation as having selected physical states that can reproduce classical values for the Casimir operators, but that are highly non-classical for the gauge-dependent operators $G_a$. 
    In contrast with this, Eq.~\eqref{eq:physcondUV} allows us to select physical states that have classical expectation values also for gauge-dependent operators.
    However, since we are only interested in gauge-invariant observables, this difference is not physically relevant. 
    In fact, we will now show that the two approaches give the same expectation values for gauge-invariant quantities.
    In other words, Eq.~\eqref{eq:physcondorbit} is consistent with the classical picture of Sec.~\ref{sec:orbitBRST}, and reflects at the quantum level the fact that the individual $G_a$ drop out of the orbit-BRST quotient of gauge-invariant observables.

    \subsubsection{Path integral and expectation values}\label{sec:pathintegralYM}

    To better understand the relation between the two approaches of Eq.~\eqref{eq:Qzerorho} and Eq.~\eqref{eq:Qzeroorbit} and the difference in the results discussed above, Eq.~\eqref{eq:physcondUV} and Eq.~\eqref{eq:physcondorbit}, it is useful to consider the expectation values on physical states as computed using the path integral. 
    In both approaches, the gauge fixing will restrict the path integral to the configurations that are compatible with the initial conditions selected by the BRST operator $Q$. The first approach, Eq.~\eqref{eq:BRSTrhocanon}, is constructed by introducing the new variables $\tilde{\rho}_a$ in such a way to cancel the gauge dependence of the constraint functions $\tilde{\phi}_{1a}=G_a-\tilde{\rho}_a$. In this way, $Q$ selects the initial conditions $G_a-\tilde{\rho}_a(\vec{x})\approx0$. As we show, we can gauge-fix the dynamics in such a way that the expectation values will reflect the gauge-dependent initial conditions $\rho_a$. The second approach, Eq.~\eqref{eq:orbitBRST}, selects the gauge invariant initial conditions of the constraint orbit, Eq.~\eqref{eq:constrOrb}. This region is the union of all gauge-equivalent constraint surfaces.

    Consider a physical state $\ket{\psi_\text{cs}}$ built following the constraint-surface approach, \textit{i.e.}, satisfying Eq.~\eqref{eq:Qzerorho}. Expectation values on this state can be written in terms of a path integral, {\it e.g.}, the expectation value of operator $F$ is:
    \begin{align}\begin{split}
       \!\!\!\! \bra{\psi_\text{cs}} F\ket{\psi_\text{cs}}=\!\!\int\!\!\mathcal{D}\tilde{\rho}_a\mathcal{D}A_{a\mu}&\mathcal{D}\Pi_{a\mu}\mathcal{D}c_a\mathcal{D}\bar{c}_a\mathcal{D}\mathcal{P}_a\mathcal{D}\bar{\mathcal{P}}_a{\mathcal{D}b_a\mathcal{D}u_a\mathcal{D}w_a}\;\\&\times F\;\exp\Bigg(\!i\!\!\int\!\! d^4x\Big(\dot{A}_{a\mu}\Pi_{a}^\mu+\dot{c}_a\mathcal{P}_a+\dot{\bar{\mathcal{P}}}_a\bar{c}_a{+\dot{u}_aw_a}\Big)-i\!\!\int\!\! dt \;H_K\Bigg),\end{split}
    \end{align}
    with all the variables on the right-hand side taken to be classical fields, rather than operators. To gauge-fix the dynamics, we compute $H_K=H+\{K,Q\}$ with $Q$ the BRST operator used in Eq.~\eqref{eq:Qzerorho} and $K$ the so-called gauge-fixing fermion. As an example, we can choose:
    \begin{align}
        K=\int d^3x\Bigg((\tilde{\rho}_a-\rho_a){w_a}+A_{a0}\mathcal{P}_a\Bigg)\;,
    \end{align}
    with $\rho_a$ functions having zero covariant derivative: $\partial_t\rho_a+C_{abc}A_{b0}\rho_c=0$, as discussed in Sec.~\ref{sec:gaugefixing}. This means that the Casimir combinations of the $\rho_a$ will be constant, which must be the case if we want to gauge-fix $\tilde{\rho}_a$ and $G_a$ to the values $\rho_a$. {In general, one will add terms proportional to $\bar{c}_a$ to $K$ in order to gauge-fix the longitudinal parts of $A_{ai}$. We omit such terms for simplicity.}
    Upon integrating out the momenta $\Pi_{ai}\,,\,\mathcal{P}_a\,,\,\bar{\mathcal{P}}_a\,,\, w_a$, this choice leads to a gauge-fixed action of the form:
    \begin{align}
        S_K=\int d^4x\Bigg(-\frac{1}{4}F_{a\mu\nu}F^{\mu\nu}_a-A_{a0}\tilde{\rho}_a+{b_a}(\tilde{\rho}_a-\rho_a)+\mathcal{L}_\text{ghost}\Bigg)\;,
    \end{align}
    where $\mathcal{L}_\text{ghost}$ depends on the ghost fields {and on $\Pi_{a0}$, but not on the $b_a$. The dependence on $\Pi_{a0}$ is usually kept in order to parameterize the gauge-fixing of the $A_{a\mu}$, as in Sec.~\ref{sec:pathintQED}}.
    Then, integrating over the {$b_{a}$} produces a Dirac delta in $\tilde{\rho}-\rho_a$, which leads to:
    \begin{align}
       \begin{split} \bra{\psi_\text{cs}} F\ket{\psi_\text{cs}}=\!\!\int_{\tilde{\rho}_a=\rho_a}\!\!\!\!\!\!\!\!\!\!\mathcal{D}A_{a\mu}&\mathcal{D}\Pi_{a{0}}\mathcal{D}c_a\mathcal{D}\bar{c}_a{\mathcal{D}u_a}\;\\&\times F\;\exp\Bigg(i\int d^4x\Big(-\frac{1}{4}F_{a\mu\nu}F^{\mu\nu}_a-A_{a0}{\rho}_a+\mathcal{L}_\text{ghost}\Big)\Bigg)\;.\end{split}
    \end{align}
    This means that expectation values can be computed fixing $\tilde{\rho}_a=\rho_a$, leading to:
    \begin{align}
        \bra{\psi_\text{cs}} G_a\ket{\psi_\text{cs}}=\rho_a\,,
    \end{align}
    when the state is normalized to one. In the path integral, a field redefinition corresponding to a BRST transformation will leave both the measure as well as the gauge-fixed action invariant, up to a redefinition of $\rho_a$ corresponding to the BRST transformation of the $\tilde{\rho}_a$: at the level of operators, $[\tilde{\rho}_a,Q]=C_{abc}\eta_b\tilde{\rho}_c$. Such transformation leaves the Casimir combinations of the $\rho_a$ invariant, e.g. $\rho_a^2=\rho^2(\vec{x})$, $d_{abc}\rho_a\rho_b\rho_c=\rho_{(3)}^3(\vec{x})$. Therefore, if $F$ is BRST-invariant, we can re-express its expectation value as an integral over $\rho_a$ at fixed Casimir combinations, with an appropriate normalization factor $\mathcal{N}$:
    \begin{align}\label{eq:averaged}
        \bra{\psi_\text{cs}} F\ket{\psi_\text{cs}}=\frac{1}{\mathcal{N}}\int_{\rho_a^2=\rho^2(\vec{x})\,,\,\dots}\!\!\!\!\!\!\!\!\!\!\!\!\!\!\!\!\!\!\!\!\!\!\mathcal{D}\rho_b\bra{\psi_\text{cs}} F\ket{\psi_\text{cs}}\;,
    \end{align}
    where the dots indicate higher Casimir combinations.
    While this redefinition leaves the matrix elements of BRST-invariant operators unchanged, it will set to zero the expectation value of the gauge-dependent, non BRST-invariant $G_a$:
    \begin{align}
        \frac{1}{\mathcal{N}}\int_{\rho_a^2=\rho^2(\vec{x})\,,\,\dots}\!\!\!\!\!\!\!\!\!\!\!\!\!\!\!\!\!\!\!\!\mathcal{D}\rho_b\bra{\psi_\text{cs}} G_a\ket{\psi_\text{cs}}=\frac{1}{\mathcal{N}}\int_{\rho_a^2=\rho^2(\vec{x})\,,\,\dots}\!\!\!\!\!\!\!\!\!\!\!\!\!\!\!\!\!\!\!\!\mathcal{D}\rho_b \;\rho_a=0\;.
    \end{align}
    Here we have used that the integral vanishes as the integration domain is gauge invariant while the integrand is gauge dependent. Indeed, if the integral were nonzero, its value would necessarily depend on the gauge. However, any gauge transformation of the integrand can be absorbed into a change of integration variables, implying that the result must be gauge invariant, and thus zero.
    This clarifies that the results of Eq.~\eqref{eq:physcondUV} and Eq.~\eqref{eq:physcondorbit} are physically equivalent. 
    In fact, the expectation values of gauge-invariant operators found with the second approach are the same as those defined on the right-hand side of Eq.~\eqref{eq:averaged}. While the results are equivalent, as we have pointed out, the second approach of the constraint orbit has the benefit of not relying on assumptions about the UV of the theory.

    Before closing this section, we discuss a further possibility that seems interesting: there could be physical states $\bra{\psi'}\,,\,\ket{\psi}$ over which the ghost combination $C_{abc}c_b\mathcal{P}_c$ has non-zero matrix elements. Using e.g. Eq.~\eqref{eq:Qzeroorbit}, in these states one might find:
    \begin{align}\label{eq:ghoststate}
         \langle \psi'|G_a|\psi\rangle=\langle \psi'|C_{abc}c_b\mathcal{P}_c|\psi\rangle=\rho_a(\vec{x})\;.
    \end{align}
    These states should have non-vanishing ghost number, although the number of Casimir ghosts may still be zero, allowing for a resolution of the constraint orbit, as discussed in Sec.~\ref{sec:gauge} and App.~\ref{app:resolution}. Moreover, since the operators $G_a$ and $C_{abc}c_b\mathcal{P}_c$ have zero ghost number, the two states must have opposite ghost number $n_g$. This is because states with non-vanishing ghost number have non-zero scalar product only with states with opposite ghost number, see e.g. \cite{Kugo:1979gm}. 
    If such states existed, then one could have states in the second approach in which the $G_a$ have expectation value $\rho_a$. This could be considered as a gauge-fixing of the ghost fields, and would show explicitly the relation between the results of Eq.~\eqref{eq:physcondUV} and Eq.~\eqref{eq:physcondorbit}.
    The existence of pairs of physical states with conjugate ghost number, $(\ket{n_g},\ket{-n_g})$ has long been debated, see \cite{Kugo:1977zq,Kugo:1979gm,Nakanishi:1995sd,Nakanishi:1996he}. For instance, in Sec.~\ref{sec:quartet} we have shown that no such state appears in QED, as every state with non-zero ghost number is BRST-exact, and therefore not physical. In general, in gauge theory, these states seem to be incompatible with either common perturbation theory or the cyclicity of the Lorentz-invariant vacuum \cite{Nakanishi:1979uk,Nakanishi:1996he}. Also, since their norm is vanishing, such states would give rise to negative-norm states, if arbitrary superpositions are allowed:
    \begin{align}
        (\bra{n_g}-\bra{-n_g})(\ket{n_g}-\ket{-n_g})<0\;.
    \end{align}
    Therefore, if Eq.~\eqref{eq:ghoststate} was a possibility, one would need to be careful in defining admissible physical states. Nonetheless, we find the existence of such states an intriguing possibility, which could be treated consistently if Eq.~\eqref{eq:ghoststate} were taken as a gauge-fixing condition.

    In conclusion, the results presented here show that the two approaches of quantizing non-abelian shadow charges are physically equivalent, and related by a gauge-fixing choice on the initial condition.

    \subsubsection{Non-abelian shadow charges are integer-valued functions}\label{sec:quantizedrho}
    Regardless of the choice of BRST construction and of physical conditions imposed on the states, we can ask whether the non-abelian nature of the Gauss' law operators restricts the eigenvalues that the Casimir operators can have. In fact, in analogy with the case of angular momentum, one can expect the shadow charges to be quantized and the Casimir operators to have eigenvalues that are integer-valued functions. This is in contrast with the case of abelian shadow charges, in which the Gauss' law operator can take arbitrary values. In order to explore this question, it is useful to construct $SU(N)$ charge operators that satisfy the algebra of the group. These can be constructed as integrals of the Gauss' law operators:
    \begin{align}\label{eq:qa}
        q_a(f)=\int d^3\vec{x} f(\vec{{x}})G_a(\vec{x})\;,
    \end{align}
    where $f(\vec{x})$ is a test function. For example, we can take $f(\vec{x})$ to be a step function with support inside of a ball of radius $R$ centered on a point $\vec{x}_0$: $f(\vec{x})=\theta(R-|\vec{x}-\vec{x}_0|)$, with $\theta(y)$ being the Heaviside theta function. Then, given the distributional commutation relations between the Gauss' law operators, analogous to Eq.~\eqref{eq:gausscommut}, we obtain:
    \begin{align}
        [q_a(f),q_b(f)]=\int d^3\vec{x} \,d^3\vec{y}\,C_{abc}G_c(\vec{x})\delta^3(\vec{x}-\vec{y})f(\vec{x})f(\vec{y})=C_{abc}q_c(f)\;,
    \end{align}
    where we have used that the step function coincides with its own square. This result is relevant, as we can now see that the square of these charges should take discrete values. In fact, suppose we have $q_a^2(f)\ket{\psi}=q^2(\vec{x}_0,R)\ket{\psi}$, with $q^2(\vec{x}_0,R)$ positive and real. Then, we can proceed as in the familiar case of angular momentum, \textit{i.e.}, $SU(2)$, in which one can find simultaneous eigenstates for $q_a^2$ and for one component, {\it e.g.}, $q_3$. Those eigenvalues must be quantized as $q_a^2|j,m\rangle=j(j+1)$,  $q_3|j,m\rangle=m$, with $-j\leq m\leq j$ and $2j\in \mathbb{Z}$.
    This has to hold for every choice of $\vec{x}_0$ and $R$ in the step function $f$. Due to additivity of the integrals, it is enough to ask that the quantization condition is satisfied at every $\vec{x}_0$ for the smallest $R$ that is resolvable in the theory, $R\sim \ell_{UV}$. Since $q_a^2$ and $q_3$ commute with $G_a^2$, we can assume the state to be an eigenstate of $G_a^2$ with eigenvalue $\rho^2(\vec{x})$. Then, for $R$ small enough with respect to the gradient of $\rho(\vec{x})$, it will also be:
    \begin{align}
        q^2(\vec{x}_0,R)=\alpha_R\;\rho^2(\vec{x}_0)R^6\;,
    \end{align}
    with $\alpha_R$ approximating $(\frac{4\pi}{3})^2$ when $R$ is smaller than the typical gradients of the fields. From this, we conclude that the quantity $\rho^2(\vec{x})\ell_{UV}^6$ must be quantized in the case of $SU(2)$:
    \begin{align}
       \alpha_{\ell_{UV}} \rho^2(\vec{x})\ell_{UV}^6=j(j+1)\;.
    \end{align}
    This means that $SU(2)$ shadow charge density in the quantum theory cannot be arbitrarily small. In practice, while $\ell_{UV}$ depends on the UV details of the theory, pair production will screen the shadow charge at scales parametrically close to the Compton wavelength of the lightest charged particle in the theory. If the theory has no charged matter fields, then the Casimir operators have eigenvalues that are integer-valued functions.

    A similar argument will hold in the case of $SU(N)$, in which one can identify $N-1$ combinations of the Gauss' law operators that are simultaneously diagonalizable, \textit{i.e.}, the Cartan subalgebra, and build lowering and raising ladder operators out of the remaining ones. As a difference with respect to the case of $SU(2)$, these ladder operators will raise or lower the diagonalized combinations of Gauss' law operators along various directions, a.k.a. the root vectors, which can be read by the commutation relations between the various combinations of operators. Still, the existence of maximum weight states that annihilate all of the raising (or all of the lowering) operators, will grant that the Casimir operators take discrete values.

    From this discussion, we conclude that non-abelian shadow charges are quantized and tend to be point-like.

 \section{Conclusions}\label{sec:conclusions}

 In this work, we have shown that considering the whole phase space of a local gauge theory exposes new sectors of infrared data, which we call shadow charges. We described these configurations in both abelian and non-abelian gauge theory, and demonstrated that their description can be made fully gauge invariant. At the classical level, we described shadow charges as deformations of the Gauss' law constraints, in Sec.~\ref{sec:classical}, and we developed a consistent way to gauge-fix the theory and describe the algebra of the observables, in Sec.~\ref{sec:gauge}. We then extended our approach and results to the quantum theory, where we analyzed the structure of physical states, and the spectrum of observables, in Sec.~\ref{sec:quantum}.
 
 From this analysis, we learned several lessons. Shadow charges can be consistently described within the pure gauge theory, without the need to introduce sources of infinite inertial mass, which would be problematic in the UV. In other words, second-class constraints directly correspond to gauge symmetry on the constraint orbit, without the need to alter the theory. This clarifies the important point that every local conservation law should correspond to a gauge symmetry. Furthermore, the BRST formalism can be extended to treat shadow charges and second-class constraints in a direct way, leading to a gauge-invariant quantized theory. Physical states are characterized by the expectation value of gauge invariant quantities, e.g. the value of the Casimir invariants built out of the Gauss' law operators. Instead, the gauge-dependent Gauss' law operators exhibit non-trivial correlations that depend on the choice of quantization. Their expectation value will vanish in the gauge-invariant quantization, while it will be non-zero if the quantization imposes a gauge-dependent expectation value. Importantly, the spectrum of shadow charges is continuous in the case of $U(1)$ but quantized for Yang–Mills theory, where the eigenvalues of the Casimir take a discrete set of values. The quantization of their spectrum is in principle controlled by a UV length-scale, but dynamically set to the Compton wavelength of the lightest charged particle in the theory.
 
 These results highlight several conceptual implications. Lorentz symmetry is broken, but only softly: the spurion $\rho({\vec x})$ has dimension three, so the renormalizable action of gauge theory is unchanged in the IR. Moreover, the theory predicts observers with arbitrary boosts with respect to the shadow charge density, suggesting that the breaking of Lorentz boosts may have no physical consequences in this case. If one interprets shadow charges as UV degrees of freedom with infinite inertial mass, it is unclear whether such a UV completion exists and whether it can be coupled consistently to gravity. Our approach sidesteps these issues by providing a self-contained IR description that does not rely on assumptions about the UV of the theory. In addition to these points, the fact that shadow charges in Yang–Mills are quantized, with values in principle tied to a UV scale, is a particularly striking feature that merits further study.
    
 Finally, these findings open the door to phenomenological applications. Shadow charges represent a new, gauge-invariant class of IR data, with possible signatures in confining dynamics, in the spectrum of gauge theories coupled to matter, and in cosmological or gravitational contexts. More broadly, our results show that only by relaxing Lorentz invariance can one describe the full local phase space of gauge theory, leading to a spectrum of IR observables significantly richer than previously assumed. From this point of view, our results suggest that full Lorentz symmetry may be an accidental feature of gauge theory, which appears in the absence of shadow charges. As a matter of fact, our results only rely on symmetry under translations and rotations, as well as the absence of dynamics for longitudinal modes.
 Exploring the full implications of this perspective is an intriguing direction for future work and may ultimately lead to assigning a dynamics to the shadow charges.
 A natural next step is to extend our analysis to general relativity, developing a full phase-space treatment of ``shadow matter" in gravity, building on Refs.~\cite{Kaplan:2023wyw,DelGrosso:2024gnj}.

	\section{Acknowledgments}
	We thank T. Cohen, R. Contino, T. Jacobson, T. Melia, and R. Rattazzi for useful discussions. We thank S. Rajendran for sharing his intimate knowledge of both cohomology and topology, as well as his encyclopedic view of higher categorical Grothendieck constructions.
	This work was supported by the U.S. Department of Energy (DOE), Office of Science, National
	Quantum Information Science Research Centers, Superconducting Quantum Materials and Systems Center (SQMS) under Contract No. DE-AC02-07CH11359. D.E.K. is supported in part by the
	U.S. National Science Foundation (NSF) under Grant No. PHY-1818899, and by the Simons Investigator Award no.~144924. L.D.G. is supported by NSF Grants No. PHY-2207502, AST-2307146, PHY-090003 and PHY-20043, by NASA Grant No. 21-ATP21-0010, by the John Templeton Foundation Grant 62840, and by the Simons Investigator Grant No. 144924. 
	
	\appendix
	\section{Dirac brackets and the rigid body}\label{app:rigidbody}
	To showcase how second-class constraints are usually treated, we find useful to consider an elementary case, the description of the rigid body as a system with constrained angular momentum.
	Consider the following Lagrangian:
	\begin{align}
		\mathcal{L}=\sum_{body}\frac{1}{2}m_i \dot{x}_i^2=\frac{1}{2}I_{ij}\dot{\theta}_i\dot{\theta}_j\;.
	\end{align}
	We can build the Hamiltonian in terms of the canonical conjugate to the Euler's angles $\theta_i$, the angular momentum components $J_i$. Having standard commutation relations between the $x_i$ and their conjugate momenta, leads to the usual algebra of rotations $\{J_i,J_j\}=\epsilon_{ijk}J_k$. In terms of these variables, the Hamiltonian is:
	\begin{align}
		H=\frac{1}{2}I^{-1}_{ij}J_iJ_j\;.
	\end{align}
	We have the following Hamilton's equations:
	\begin{align}
		\dot{J}_i=\{J_i,H\}=\frac{1}{2}I^{-1}_{lm}\{J_i,J_lJ_m\}=\epsilon_{ijk}I^{-1}_{jl}J_jJ_k\;,
	\end{align}
	which describe the precession of angular momentum.
	We can see that the total angular momentum $J^2=\sum_i J_i^2$ is constant: $\{J^2,H\}=0$.
	We now want to describe the system in terms of constraints. 
	The simplest configurations are those of zero angular momentum, with constraints $\phi_{0i}=J_i\approx 0$. These constraints are first class, $\{\phi_{0i},\phi_{0j}\}\approx 0$, and do not imply any secondary constraints, as their time derivatives automatically vanish on the constraint surface. 
	
	As a more interesting case, we could consider shifted constraints of the form $\phi_{0i}=J_i-l_i\approx 0$. These constraints will not be first class for non-vanishing $l_i$: $\{\phi_{0i},\phi_{0j}\}=\epsilon_{ijk}(\phi_{0k}+l_k)\approx \epsilon_{ijk}l_k$. Among these, the Casimir combination $\phi_J=l_i\phi_{0i}$ will be first class, as $\{\phi_J,\phi_{0j}\}=l_i\{J_i,J_j\}=0$.
	Time evolution of these constraints would imply the following secondary constraints:
	\begin{align}
		\phi_{1i}=\dot{\phi}_{0i}=\epsilon_{ijk}I^{-1}_{jn}J_jJ_n-\partial_t l_i\approx\epsilon_{ijk}I^{-1}_{jn}l_jl_n-\partial_t l_i\approx 0\;.
	\end{align}
	These describe the precession of the constrained values, as dictated by the equations of motion. Clearly, the fact that the Hamiltonian depends on the $J_i$ makes it not very significant to constrain these variables when they have non-zero values, as their time evolution is non-trivial, resulting in the non-trivial secondary constraints above. In other words, the system can be sensitive to rotations (generated by the $J_i$, and these should not necessarily be treated as symmetries. In any case, we find that this example is close enough to the case of Yang-Mills shadow charges to make this discussion useful.
	
	In the specific case in which the angular momentum is aligned along one of the principal axes of the body -- $\hat{x}$ -- we can rotate our basis in such a way that $I$ is diagonal and get: 
	\begin{align}
		\phi_{0x}=J_x-l_x&\approx 0\;,\;\phi_{0y}=J_y\approx 0\;,\;\phi_{0z}=J_z\approx 0\;,\\
		\{\phi_{0x},\phi_{0y/z}\}&\approx 0\;,\;\{\phi_{0y},\phi_{0z}\}\approx l_x\;.
	\end{align}
	Chosing the other principal axes along $\hat{y}$ and $\hat{z}$, we will have that no other constraints appear, besides $\partial_t l_x\approx 0$\;.
	
	Given these commutation relations, a common approach to describe the system is introducing Dirac brackets, as explained around Eq.~\eqref{eq:diracbrack}. This approach consists in by-hand demoting the second-class constraints to algebraically trivial functions, whose Dirac brackets with any other quantity are vanishing.
	Here we describe this approach, outlining its dynamical meaning. 
	
	Dirac brackets are defined in terms of the Poisson brackets as: 
	\begin{align}
		\{A,B\}_D=\{A,B\}-\{A,\phi_\alpha\}\mathcal{B}^{-1}_{\alpha\beta}\{\phi_\beta,B\}
	\end{align}
	here $\{\phi_\alpha,\phi_\beta\}\approx \mathcal{B}_{\alpha\beta}$ and the inverse is defined in the subspace in which $\mathcal{B}_{\alpha\beta}$ is invertible.
	In our case, this subspace is given by $\phi_{0y}$ and $\phi_{0z}$, with $\mathcal{B}_{yz}=l_x$. By construction, the Dirac brackets of first class constraints are equal to the Poisson brackets, while the Dirac brackets of a second class constraint vanish with every quantity:
	\begin{align}
		\{\phi_{0y},A\}_D=\{\phi_{0y},A\}-\{\phi_{0y},\phi_{0a}\}C^{-1}_{ya}\{\phi_{0a},A\}=\{\phi_{0y},A\}-\delta_{ya}\{\phi_{0a},A\}\equiv 0\;.
	\end{align}
	Since $\phi_{0y}=J_y-l_y$ is now algebraically trivial, it means that the Dirac brackets are making $J_y$ algebraically trivial. This means that Dirac brackets, in the case at hand, subtract the precession effects from the evolution of the system. Therefore, the vanishing of Dirac brackets, e.g. of a given quantity with the Hamiltonian, indicates conservation of the given quantity up to precession effects.
	
	As we discuss in Sec.~\ref{sec:QYM}, the use of Dirac brackets in quantum field theory is in many cases impractical due to the difficulty in finding operators whose commutators give a representation of the algebra of Dirac brackets.
	Instead, as we discuss in Sec.~\ref{sec:UVfields} and~\ref{sec:constraintorbit}, second class constraints can be treated by either introducing new physical variables and building new constraints, or by studying the system on the constraint orbit.
	
	\section{Shadow charges from infinite-mass fields}\label{sec:rhocanonical}	
	In the following, we discuss in some more detail the approach described in Sec.~\ref{sec:UVfields} to describe systems with second-class constraints. In this approach the phase space and its Poisson bracket structure is extended in such a way to make all of the constraints of Eq.~\eqref{eq:constrYM} first class.
    Again, we stress that retrieving the gauge symmetry associated to these constraints is relevant, as it allows to consistently set aside the non-dynamical and non-physical variables of the theory.
	
	The way we do it follows the simple idea that a pair of second class constraints can always be seen as a combination of a pair of first class constraint subject to certain gauge fixing conditions, provided the phase space is suitably extended \cite{Henneaux:1992ig,Batalin:1986aq,Batalin:1986fm,Batalin:1991jm}. As an example, consider a system of two conjugate canonical variables, $q_1\,,\,p_1$ subject to the constraints: $q_1\approx 0\,,\,p_1\approx 0$. Due to canonical Poisson brackets between $q$ and $p$, these constraints are second class. However, the same situation can be described by adding two new conjugate variables, $q_2\,,\,p_2$, and changing the constraints to:
	\begin{align}
		q_1+q_2\approx0\;,\quad p_1-p_2\approx0\;,
	\end{align}
	which are first class and reduce to the original constraint surface once one fixes the new variables to $q_2\,,\,p_2\approx 0$.
	
	In the case of Yang-Mills theory with shadow charges, one can extend the phase space by introducing $n$ new canonical variables indexed by the adjoint index $a$, which we call $\tilde{\rho}_a$, representing the shadow charges. In this setup, the algebra of Poisson brackets can be extended to this new portion of phase space by defining new Poisson brackets as in Eq.~\eqref{eq:rhocanon}:
	$$
		\{\tilde{\rho}_a,\tilde{\rho}_b\}\equiv C_{abc}\tilde{\rho}_c\;,
	$$
	where we are omitting a delta function in position space, and we are defining Poisson brackets evaluated on products of $\tilde{\rho}_a$s in such a way that they satisfy the Leibniz rule.
	Since the Hamiltonian is independent on the new variables $\tilde{\rho}_a$, we can describe the constraint surface in Eq.~\eqref{eq:constrYM} in terms of:
	\begin{align}\label{eq:constrsurfextra}
		\phi_{0a}=\Pi_{a0}\approx0\;,\quad	\phi_{1a}=G_a-\tilde{\rho}_a\approx0\;.
	\end{align}
	Now, due to the non-trivial Poisson brackets between the $\tilde{\rho}_a$, we see that the constraints $\phi_{1a}$ become first class:
	\begin{align}
		\{\phi_{1a},\phi_{1b}\}\approx C_{abc}\phi_{1c}\approx 0\;.\quad
	\end{align}
	The strategy of introducing a Poisson bracket structure for the shadow charges has a very clear physical interpretation. Indeed, it corresponds to identifying the $\rho_a$ with charge densities of a non-dynamical background charged field. 
	
	To show how this is the case, one can note that in a theory for gluons coupled to a field charged under $SU(N)$, the Gauss' law operators are:
	\begin{align}
		\tilde{G}_a=G_a-J_{a}^0\;,
	\end{align}
	with $J_a^0$ the time component of the matter field's $SU(N)$ Noether current. The Gauss' law operators in the theory with matter have the same Poisson brackets as the $G_a$:
	\begin{align}
		\{\tilde{G}_a,\tilde{G}_b\}=C_{abc}\tilde{G}_c\;,
	\end{align}
	and they are conserved, $\{\tilde{G}_a,\tilde{H}\}=0\;$, where $\tilde{H}$ is the Hamiltonian of the system coupled to matter, with no dependence on $A_{a0}$.
	This means that the new Gauss' law operators $\tilde{G}_a$ will be first class constraints on the following constraint surface:
	\begin{align}
		\Pi_{a0}\approx 0\;,\quad\tilde{G}_a\approx 0\;.
	\end{align}
	Taken this constraint surface, one can then focus on the restricted phase space where the matter field variables are set to values that entirely fix the charge densities, $J_a^0\equiv \rho_a$, completely removing them from the dynamics.
	This reduction of the dynamics will be compatible with the dynamics of the new theory only when the matter field's inertial mass is formally infinite. Indeed, this limit implies that the matter field's charge density must be conserved independently of the gluon, since no finite amount of momentum transferred to the matter field can move its density profile. This implies the conservation of the $G_a$ as prescribed by $H$ in the original phase space.
	This limit therefore reproduces the constraint surface of Eq.~\eqref{eq:constrYM}.
	As we have shown above, one can skip the steps of introducing unmovable background matter fields by only introducing canonical variables corresponding to the charge densities and their respective Poisson brackets. However, it is clear that the modification of Poisson brackets in Eq.~\eqref{eq:rhocanon} is inherited by the infinite inertial mass fields. Therefore, it seems that the approach presented here is subtly tied to a non-standard UV completion of the theory.

    Such a UV completion, as we have commented, seems to be non trivial, as the Hamiltonian of these fields might be ill-defined. Moreover, such fields would be in tension with gravity, as one would either have to deal with an ill-defined gravitational field, or to make the gravitational mass of these fields finite, which might be irreconciliable with invariance under diffeomorphisms.
	
	With this, we have proved that it is possible to add physical variables in such a way that no generator of gauge symmetry is broken, regardless of the shadow charge. 
	As we have discussed in the rest of the text, this procedure changes the physics of the theory, making it consistent to predict configurations beyond the shadow charges that the are predicted by the original Yang-Mills theory.
	In comparison, the result discussed in Sec.~\ref{sec:constraintorbit} shows that gauge symmetry is not affected by the presence of shadow charges, even when one does not change the physical content of the theory. Therefore, shadow charges can be consistently thought of as features of the gauge theory without the need to invoke new charged fundamental fields.

    \section{Resolving the constraint surface in an explicit example}\label{app:resolution}
    Here we want to illustrate how the definition of the differential $\delta$ given in Eq.~\eqref{eq:deltadef} allows to resolve the constraint surface algebraically, as in Eq.~\eqref{eq:homology}.
    We use an explicit example of a phase space of four real variables $x_i$, $i=1,\dots,4$, and two constraints $\phi_1=x_1-x_3\,,\,\phi_2=x_2-x_4$. We can represent explicitly the differential $\delta$ of Eq.~\eqref{eq:deltadef} that resolves the algebra of functions on $\phi_{1}\approx 0\,,\,\phi_2\approx 0$ by introducing two new variables $z_1\,,\,z_2$, \textit{i.e.}, the ghost momenta, and defining:
	\begin{align}\label{eq:deltaex}
		\delta \,g(x,z)=\Big(\phi_1\frac{\partial^L}{\partial z_1}+\phi_2\frac{\partial^L}{\partial z_2}\Big)g(x,z)\;.
	\end{align}
    Here, we have defined the left derivative of a function $\frac{\partial^L}{\partial z}g$ as defined by variation of $g$ written with the variation of $z$ on the left, $g(z+\Delta z)\simeq g(z)+\Delta z\, \frac{\partial^L}{\partial z}g$. This makes it possible to take $z$ to be fermionic without ambiguities.
    
    Regardless of whether $z$ are fermionic or bosonic, this definition grants that the algebra of the original phase space is contained in the Kernel of $\delta$, as $\delta f(x)=0$. However, we can see that the Kernel might have terms that depend on the ghost momenta $z$, e.g. functions that depend on $z_1\phi_2-z_2\phi_1$.
    We can isolate the original phase space in the Kernel of $\delta$ by exploiting the grading defined by the ghost number. In fact, if we consider the zero ghost number part of the Kernel of $\delta$, $\text{Ker}(\delta)_0$ we obtain exactly the functions on the original phase space, $f(x)$. Similarly, the zero ghost number part of the image of $\delta$, $\text{Im}(\delta)_0$, will be proportionalto $\phi_1$ and $\phi_2$, with no ghost terms. Therefore, the zero ghost number quotient of the Kernel of $\delta$ by its image, will give the reduced phase space of functions on the constraint surface
    $\phi_1\approx\phi_2\approx 0$:
    \begin{align}
        \frac{\text{Ker}(\delta)_0}{\text{Im}(\delta)_0}=C^\infty(\{\phi_i\approx0\})\;.
    \end{align}
    
    While this does not depend on the commutativity of the ghost momenta $z$, we can see that taking the ghost momenta to be Grassmannian makes $\delta$ a differential, $\delta^2=0$. In addition to this, the Grassmannian nature of the ghost momenta grants that at non-zero ghost number, all the quantities in the Kernel of $\delta$ are also in the image of $\delta$. This means that the quotients of Kernel by image of $\delta$ at non-zero ghost number are vanishing.
    To show these points in our example, note that when the $z$ are Grassmannian, the functions defined over the extended phase space take the form:
    \begin{align}
        g(x,z)=h(x)+a_1(x)z_1+a_2(x)z_2+b(x)z_1 z_2\;.
    \end{align}
    This means that $\delta^2 g=b(x)\delta^2 z_1z_2$. However, since the $z$ are Grassmannian, the left derivatives in Eq.~\eqref{eq:deltaex} will satisfy a fermionic Leibnitz rule, \textit{i.e.}, with a minus sign rather than a plus sign. For this reason we have:
    \begin{align}
        \delta^2 z_1 z_2=\delta (\phi_1z_2-z_1\phi_2)=\phi_1\phi_2-\phi_1\phi_2=0\;,
    \end{align}
    meaning that $\delta$ is indeed a differential.
    In addition to this, we can see that the higher ghost number sectors of the Kernel and of the image of $\delta$ are the same. For instance, for ghost number 1 and 2, we have:
    \begin{align}
        \text{Ker}(\delta)_1&=a(x)(\phi_2 z_1-\phi_1z_2)=\delta(a(x)z_1 z_2)=\text{Im}(\delta)_1\;,\\
        \text{Ker}(\delta)_{2}&=\text{Im}(\delta)_2=0\;.
    \end{align}
    Instead, higher ghost number Kernels and images of $\delta$ vanish trivially. This means that taking the only quantities to be $\delta$-closed but not $\delta$-exact in the extended phase space are the functions with support on the constraint surface.
    
	\section{Non-Abelian Gauge Symmetry of the Lagrangian}\label{app:gaugesymm}
	Here we briefly show how the action in Eq.~\eqref{eq:simpleactionYM} is invariant under gauge transformations $A_{a\mu}\to A_{a\mu}+(D_\mu\lambda)_a$. 
	This invariance is ensured owing to the condition $(D_0\rho)_a\approx0$ derived in Eq.~\eqref{eq:basicallyphi2}. Indeed, when we transform the gauge field as $A_{a\mu}\to A_{a\mu}+(D_\mu\lambda)_a$, we have the following variations:
	\begin{align}\begin{split}
			&\delta_\lambda(F_{a\mu\nu}F^{\mu\nu}_a)=0\;,\\
			&\delta_\lambda(\rho_a A_{a0})= \rho_a (\dot{\lambda}_a+C_{abc}A_{b0}\lambda_c)=\frac{d}{dt}(\rho_a\lambda_a)-\lambda_a(D_0\rho)_a\approx \frac{d}{dt}(\rho_a\lambda_a)\;,\end{split}
	\end{align}
	meaning that the equations of motion do not change under these transformations.
	This result can be understood as a consequence of eliminating the conjugate momenta $\Pi_{ai}$ using the non-abelian Ampere's law, $\dot{A}_{ai}=\{A_{ai},H_K\}$.
	Indeed we have:
	\begin{align}
		(D_i\dot{\lambda})_a=&\{\dot{A}_{ai},\lambda_bG_b+u_b\Pi_{b0}\}=\{\Pi_{ai}+(D_iA_0)_{a},\lambda_bG_b+u_b\Pi_{b0}\}\;\\
		=&(D_i u)_a-C_{abc}A_{b0}(D_i\lambda)_c+C_{abc}(\dot{A}_{bi}-(D_iA_0)_b)\lambda_c
	\end{align}
	This can be further manipulated to obtain:
	\begin{align}
		\partial_i(\partial_t\lambda_a+C_{abc}A_{b0}\lambda_c)+C_{abc}A_{bi}(\partial_t\lambda_c+C_{cde}A_{d0}\lambda_e)=(D_iu)_a\;,
	\end{align}
	which is satisfied only provided $u_a=(D_0\lambda)_a$. This shows that the usual gauge transformations $A_{a\mu}\to A_{a\mu}+(D_\mu\lambda)_a$ are recovered from the gauge transformations generated by the constraints in the canonical formalism.

	\section{Boosted shadow charges and Lorentz transformations}\label{sec:boosts}
    The derivation we followed in the main text restricts our results to shadow charges that are time independent. Despite this, the fact that the action can be expressed in a Lorentz-covariant form, see e.g. Eq.~\eqref{eq:gauge_fixed_action_QED_FP}, makes it clear that the time-independence of the shadow charges is not a physical restriction of the possible configurations. 
    In fact, for each configuration in which the shadow charge can be compared to a physical observer on a given trajectory, e.g. a cow, $(\rho(\vec{x})+ \mathrm{cow})$ there exist other configurations, predicted by the Hamiltonian of the theory, in which the shadow charge is the same but the physical observer is boosted with respect to the shadow charge: $(\rho(\vec{x})+\mathrm{cow}')$. In other words, the theory predicts shadow charges with arbitrary boosts, with respect to any physical observer. 
    Nevertheless, the Lorentz group will act trivially on states that contain a shadow charge and nothing else, suggesting that Lorentz symmetry may be broken.
    
	Regardless of Lorentz symmetry, these considerations prompt us to expect that the phase space of the theory should contain configurations in which the shadow charge is boosted.
	As a matter of fact, if the phase space is taken to include $A_0$, then a Lorentz transformation can be implemented as a change of variables in phase space, mapping the phase space into itself and resulting in a description of the constrained dynamics from a boosted frame, in which the shadow charge is moving.
	As we will see, the boosted constraint surfaces will be singled out through the invariants under translations along the boosted time coordinate.
	In other words, boosted shadow charges will be invariant under boosted Hamiltonians.
	The first step, therefore, should be to derive a boosted Hamiltonian.
	Let us consider a boosted frame $O'$ with speed $\vec{v}=v\hat{v}$ relative to our reference frame $O$, which we have implicitly used so far. The space-time coordinates in $O'$ are $t'=\gamma(t-\vec{v}\!\cdot\!\vec{x})\;,\; \vec{x}'=\gamma(\vec{x}-\vec{v}t)$, with $\gamma=1/\sqrt{1-v^2}$ the usual Lorentz boost factor. 
	Thinking of the Hamiltonian as the time component of the four-momentum of the system, it is clear that we should expect the Hamiltonian of the boosted frame to take the form:
	\begin{equation}\label{eq:boostedHamiltonian}
		H_{o'} = \gamma\left(H_o - \vec{v}\cdot \vec{P}_{o}\right) \, ,
	\end{equation}
	where $H_o$ and $\vec{P}_o$ are the Hamiltonian and total momentum in our frame $O$. While the Hamiltonian is given by e.g. Eq.~\eqref{eq:H}, defining the space momentum of the system is a more subtle issue. One might think of expressing it in terms of the Poynting vector of the fields:
	\begin{equation}\label{eq:poynting}
		\vec{P}_{rad,i} = \int d^3 x \, (\vec{\Pi} \wedge\vec{B})_i\,
		=\int d^3x\,(\vec{\Pi}_j\vec{\nabla}_i\vec{A}_j-\vec{\Pi}_j\vec{\nabla}_j\vec{A}_i)\;,
	\end{equation}
	with $\vec{B} = \vec{\nabla} \wedge \vec{A}$. 
	However, it is simple to see that this quantity cannot account for the momentum carried by a moving shadow charge. Indeed, $\vec{P}_{rad}$ has vanishing Poisson brackets with the Gauss' law operator, meaning that $H_o-\vec{v}\!\cdot\!\vec{P}_{rad}$ does not describe invariant surfaces with boosted shadow charge.
	
	As a matter of fact, the Hamiltonian in the boosted frame should be the canonical generator of translations along $t'$. This means that it must be:
	\begin{align}
		\{F,H_{o'}\}=\frac{d}{dt'}F=\gamma\Big(\frac{d}{dt}-\vec{v}\!\cdot\!\frac{d}{d\vec{x}}\Big)F=\gamma(\{F,H_o\}+\vec{v}\!\cdot\!\{F,\vec{P}_o\})\;.
	\end{align}
	However, we can see that the Poynting vector does not generate the translations for the quantities $\vec{A}\,,\,\vec{\Pi}$:
	\begin{align}
		\{\vec{A}_j,\vec{P}_{rad,i}\}&=\int d^3x\{\vec{A}_j,(\vec{\Pi}\wedge\vec{B})_i\}=\epsilon_{ijk}\vec{B}_k=\vec{\nabla}_i \vec{A}_j-\vec{\nabla}_j\vec{A}_i\;,\\
		\{\vec{\Pi}_j,\vec{P}_{rad,i}\}&=\int d^3x\{\vec{\Pi}_j,(\vec{\Pi}\wedge\vec{B})_i\}=\epsilon_{i\ell k}\epsilon_{kmj}\vec{\nabla}_m\vec{\Pi}_\ell=
		\vec{\nabla}_i\vec{\Pi}_j-\delta_{ij}\vec{\nabla}\!\cdot\!\vec{\Pi}\;.
	\end{align}
	This means that the Pointing vector does not capture the whole momentum associated with the electromagnetic field's configuration. In fact, it only generates translations of the transversal fields.
	Instead, we can see that the spatial translations are generated by the following quantity:
	\begin{align}
		\vec{P}_{o,i}=\int d^3x \;\vec{\Pi}_j\vec{\nabla}_i\vec{A}_j\;.
	\end{align}
	In fact this leads to:
	\begin{align}
		\{\vec{\Pi}_j,\vec{P}_{o,i}\}=\vec{\nabla}_i\vec{\Pi}_j\;,\quad\{\vec{A}_j,\vec{P}_{o,i}\}=\vec{\nabla}_i\vec{A}_j\;.
	\end{align}
	With this, we find that the Hamiltonian of the boosted frame $H_{o'}$ has invariant surfaces corresponding to boosted shadow charges:
	\begin{align}
		0=\{\vec{\nabla}\!\cdot\!\vec{\Pi}',H_{o'}\}=\gamma^2\Big(\frac{d}{dt}\vec{\nabla}\!\cdot\!\vec{\Pi}-\vec{v}\!\cdot\!\vec{\nabla}(\vec{\nabla}\!\cdot\!\vec{\Pi})\Big)\,,
	\end{align}
	where we have used that $\vec{\nabla}\!\cdot\!\vec{\Pi}'=\gamma\vec{\nabla}\!\cdot\!(\vec{\Pi}+\vec{v}\wedge\vec{B})=\gamma\vec{\nabla}\!\cdot\!\vec{\Pi}$. This equation implies that the evolution according to $H_{o'}$ is compatible with a constraint of the form:
	\begin{align}
		\vec{\nabla}\!\cdot\!\vec{\Pi}-\rho(\vec{x}+\vec{v}t)\approx 0\;,
	\end{align}
	that is, a boosted shadow charge in our reference frame $O$.
	
	The departure from the naive expectation that the Poynting vector in Eq.~\eqref{eq:poynting} should generate translations comes from the fact that a given shadow charge breaks introduces longitudinal background fields, which commute with the Poynting vector. In terms of total momentum, one can understand the need for the extra longitudinal part from the fact that the term $J^\mu_\text{S}A_\mu$ in the action explicitly depends on the position, leading to a non-conserved stress energy tensor. The non-conservation corresponds to an implicit momentum injection in the system, needed in order to keep the shadow charge fixed.
	
	\bibliographystyle{JHEP}
	\bibliography{biblio}  
\end{document}